\documentclass{aa} 
\usepackage{graphicx}
\usepackage{txfonts}
\begin{document}
   \title{VLT-VIMOS  integral field spectroscopy of Luminous and Ultraluminous Infrared Galaxies}

   \subtitle{I. The sample and first results}

   \author{Santiago Arribas\inst{1}, Luis Colina\inst{1}, Ana Monreal-Ibero\inst{2}, 
          Julia Alfonso\inst{1}, Macarena Garc\' ia-Mar\' in\inst{1},
          \and
          Almudena Alonso-Herrero\inst{1}
          }

   \offprints{S. Arribas}

   \institute{Departamento de Astrof\'{\i}sica Molecular e Infrarroja (DAMIR),
                    Instituto de Estructura de la Materia (IEM/CSIC), C/ Serrano, 121, 28006-Madrid, Spain.\\
              \email{arribas,colina,julia,maca,aalonso@damir.iem.csic.es}
              \and
              Instituto de Astrof\'{\i}sica de Canarias,  38200-La Laguna, Tenerife, Spain.\\
              \email{amonreal@iac.es}
             }

   \date{}

 
  \abstract
   {(Ultra)Luminous Infrared Galaxies [(U)LIRGs] are much more numerous at cosmological distances than locally, and are likely the precursors of elliptical galaxies. Studies of the physical structure and kinematics of representative samples of these galaxies at low redshift are needed in order to understand the interrelated physical processes involved. Integral field spectroscopy (IFS) offers the possibility of performing  such a detailed analysis.}
   {Our goal is to carry out IFS of 42 southern systems which are part of a representative sample of about 70 low redshift (z $\le$ 0.26) (U)LIRGs, covering different morphologies from spirals to mergers over the entire infrared luminosity range (L$_{IR}$=10$^{11}-10^{12.6}$L$_{\odot}$).}
   {The present study is based on optical IFS obtained with the VIMOS instrument on the VLT.}
   {The first results of the survey are presented here with the study of two galaxies representative of the two major morphological types observed in (U)LIRGs, interacting pairs and morphologically- regular, weakly-interacting spirals, respectively. We have found that IRAS F06076$-$2139 consists of two low-intermediate mass (0.15 and 0.4 m$_*$) galaxies with relative velocities of $\sim$  550 km s$^{-1}$ and, therefore, it is unlikely that they will ever merge. The VIMOS IFS has also discovered the presence of a rapidly expanding and rotating ring of gas in the southern galaxy (G$_s$). This ring is interpreted as the result of a nearly central head-on passage of an intruder about 140 million years ago. The mass, location and relative velocity of the northern galaxy (G$_n$) rules out this galaxy as the intruder.
   IRASF 12115$-$4656 is a spiral for which we have found a mass of 1.2 m$_*$. The ionized gas shows all the kinematic characteristics of a massive (8.7 $\times$ 10$^{10}$M$_{\odot}$), fast rotating disk. The neutral gas traced by the NaI doublet shows distinct features not completely compatible with pure rotation. The neutral and ionized gas components are spatially and kinematically decoupled. The analysis presented here illustrates the full potential of IFS in two important aspects: (i) the study of the kinematics and ionization structure of complex interacting/merging systems, and (ii) the study of the kinematics of the different gas phases, neutral (cool) and ionized (warm), traced by the NaD and H$\alpha$ lines, respectively.} 
   {}


\authorrunning{(VIMOS (U)LIRGs first results}          
\maketitle
%

\section{Introduction}

   Most of the observational efforts devoted in recent years to the study of galaxy formation and evolution have been focused on obtaining and analyzing large surveys of different characteristics such as, for instance,   2MASS (Skrutskie et al. 2006), SDSS (Adelman-McCarthy et al. 2006), 2DF (Colless et al. 2001), UDF (Beckwith et al.  2006),  GOODS (Giavalisco et al. 2004),  COSMOS (Capak et al. 2007),  GEMS (Rix et al. 2004),  etc.  Thanks to these studies observational constrains on integrated properties such as  the galaxy luminosity functions, galaxy (stellar) masses,  sizes,  spectral energy distributions,  etc., as well as their evolution with redshift,  have drastically improved (see, for instance, Rix et al. 2006 and references therein).  However, the observational limits imposed by the HST and 10 meter class telescopes have been reached and a substantial new improvement using this methodology will probably have to wait until a new generation of telescopes (JWST, ELT, etc) comes into operation.
   
In addition to those integrated properties, a comprehensive picture of galaxy formation should explain the internal structure of galaxies (i.e., internal kinematics, internal stellar population gradients, dust distribution, ionization structure,  nuclear properties and effects, interaction with the IGM, etc).  Such detailed studies are more demanding in terms of 'data quality per object'  requiring high S/N two dimensional spectroscopic information with high angular and spectral resolution.
   Thanks to the advent and popularization of Integral Field Spectroscopic instruments it is now possible to obtain high quality two-dimensional spectroscopy (i.e. 3D data)
   over a wide wavelength range ($>$ 1000 $\AA$) in one spectral setting. Although most such studies have been based on local samples of galaxies (e.g.,  Mediavilla et al. 1997; Peletier et al. 1999, 2007; del Burgo et al. 2001; de Zeeuw et al. 2002 and references therein), recent work also includes high-z studies (e.g., Smith et al. 2004; Forster-Schreiber et al. 2006; Flores et al. 2006; Puech et al. 2007; Law et al. 2007). However, limitations in sensitivity and angular resolution make the studies of the internal structure of high-z galaxy populations via IFS extremely difficult, and the observation of comprehensive samples (i.e., not just individuals at the tip of the luminosity function) will require larger aperture telescopes, and higher angular resolution than those currently available.
   
   In the context of the high-z universe, the study of representative samples of some low-z galaxies populations are of particular importance. That is the case of the luminous and ultraluminous infrared galaxies  (LIRGs,  L$_{IR}$=L[8-1000$\mu$m]=10$^{11}$-10$^{12}$L$_{\odot}$, and   ULIRGs, L$_{IR}$$>$10$^{12}-10^{13}$L$_{\odot}$, respectively). These objects are thought to be the local counterparts of cosmologically important galaxy populations at z $>$ 1 such as the submillimiter galaxies (SMG; e.g., Frayer et al. 2004), and the majority of the infrared sources found in recent {\it Spitzer} cosmological surveys  (e.g., P\'erez-Gonz\'alez et al. 2005).  ULIRGs and LIRGs  have large amounts of gas and dust, and are undergoing an intense star formation in their (circum)nuclear regions (e.g. Scoville et al. 2000, Alonso-Herrero et al. 2006, and references therein). This starburst activity is believed to be their major energy reservoir, although AGN activity may also be present (Genzel et al. 1998).   In many cases (especially for ULIRGs) these objects show clear signs of an on-going merging process producing moderate-mass (0.1$-$1 m$_*$) ellipticals as the end product of the merger (Dasyra et al. 2006, Genzel et al. 2001).  Although we have a qualitative picture, details on how the physical and dynamical processes involved are interrelated and how they determine the internal structure of these cosmologically important objects are far from being well understood.  
    
    We have already started a program aimed at studying the internal structure and kinematics of a representative sample of low-redshift LIRGs and ULIRGs using optical and near-IR Integral Field Spectroscopy. These objects are natural laboratories that allow to investigate in detail the internal structure of these galaxy populations in a high S/N, high spatial resolution regime, capturing their complex two dimensional structure. We are mainly analyzing rest-frame optical and near-IR spectral diagnostic features on linear scales of about 1 kpc, that will appear in the near- and mid-IR in the high-z galaxy populations to be studied with future instruments such as the NIRSpec and MIRI instruments on board the JWST. This program initially has been focused on the most luminous objects, the ULIRGs  (e.g., Colina, Arribas \& Monreal-Ibero, 2005 and references therein; Garc\'{\i}a-Mar\'\i n 2007) and now is being extended towards lower luminosities, i.e. the LIRG luminosity range. 
    
    The present paper is the first of a series of studies based on a representative sample of LIRGs observed with VLT-VIMOS. With $\sim$ 2000 spectra per object, this study involves a large analysis effort. In this first paper we describe the sample, the experimental set-up used for all the observations with VIMOS, as well as the reduction procedures and analysis methods. The analysis of two of the galaxies representative of the sample is shown to illustrate and emphasize the power of the survey in several areas, which after completion will include, among others, the following: (i) two dimensional kinematics of different phases (neutral and ionized) of the interstellar gas as a function of luminosity and morphology, (ii) two dimensional  ionization structure, and the interplay between ionization and kinematics (i.e., the role of shocks), and (iii) star formation in the (circum)nuclear and external regions (candidate Tidal Dwarf Galaxies and/or precursors of globular clusters)  in interacting/merging systems.  Throughout the paper we will consider H$_{0}$=70 km~s~$^{-1}$Mpc$^{-1}$, $\Omega_{M}$= 0.7, $\Omega_{\Lambda}$= 0.3.


\section{The IFS (U)LIRG Survey and the VIMOS sample}

The IFS (U)LIRG Survey is a large program aimed at the study of the two dimensional internal structure of a representative sample of LIRGs and ULIRGs using integral field spectroscopic facilities both in the southern (VIMOS, Le F$\grave{e}$vre et al. 2003), and northern hemisphere  (INTEGRAL, Arribas et al. 1998 and PMAS-CAHA, Roth et al. 2005). The complete survey contains about 70 galaxies covering the entire luminous and ultraluminous infrared luminosity range, and representing the different morphologies (spirals, early interactions, advanced mergers, and post-mergers) observed in this class of objects. 

The VIMOS sample contains  a total of 42 LIRGs and ULIRGs (see Table 1 for general properties). 
The LIRG sample consists of 32 systems (42 galaxies) drawn from the IRAS Revised Bright Galaxy Sample (RBGS, Sanders et al. 2003), and contains all the southern systems from the complete volume-limited sample of LIRGs of Alonso-Herrero et al (2006).   The LIRG subsample has a mean redshift of 0.022 ($\pm$ 0.010), with distances ranging from $\sim$ 40 to 220 Mpc and covers the entire luminosity range.

   \begin{figure*}
   \centering
   \vskip -3cm
     \includegraphics[width=18cm]{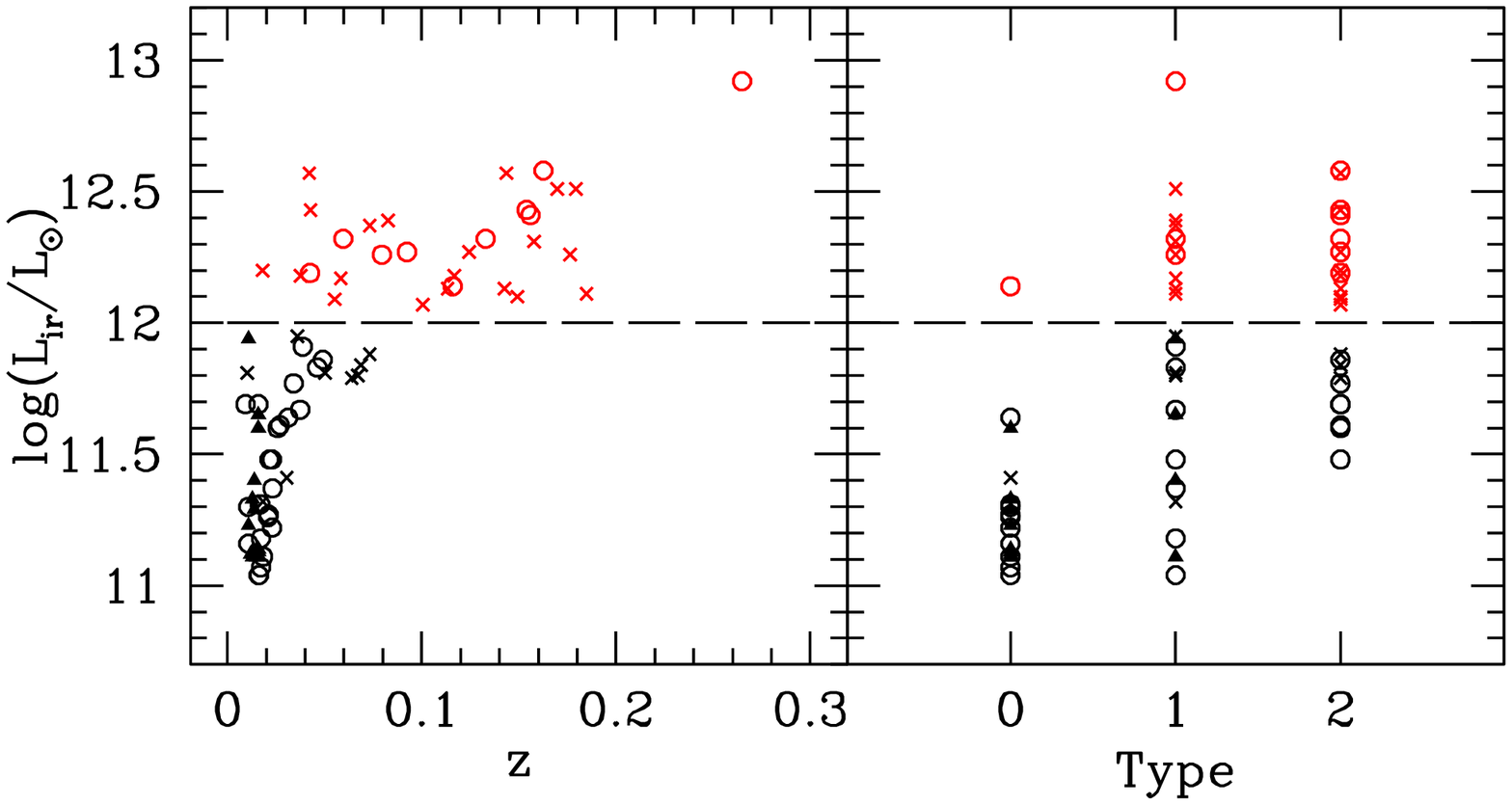}
      \caption{ Distribution of the objects of the sample in the luminosity - redshift (left) and luminosity - morphological/merging type (right) planes, where 0 indicates a single isolated object, 1 interacting pairs in a pre-coalecense phase, and 2 a single system with evidence of having suffered a merger (i.e.  post-coalescence phase).  Circles correspond to the current VIMOS sample, whereas crosses and filled triangles are the INTEGRAL and PMAS samples, respectively (see text). The dashed line divides the sample in LIRGs (black symbols) and ULIRGs (red symbols).       }
         \label{Fig1}
          \end{figure*}

 The ULIRG sample consists of 10 systems (13 galaxies) selected from the IRAS 1 Jy sample of ULIRGs (Kim et al.  2002), the RBGS (Sanders et al. 2003), and some objects from the HST/WFPC2 snapshot sample (ID 6346 PI: K. Borne), with an average redshift of 0.126 ($\pm$0.064).

Galaxies in the VIMOS sample were selected such as to cover not only the relevant luminosity range but also representing the different morphological types identified in these systems. The infrared luminosity, L$_{ir}$=L(8-1000$\mu$m), 
was calculated using the fluxes in the four IRAS bands according to the standard prescription (see Table 1 in Sanders \& Mirabel 1996). The values obtained are slightly different from those given by Sanders et al. (2003) due to the different luminosity distances considered (typically our values are $\sim$ 20\% higher than those by  Sanders et al.). Since the definition of LIRGs and ULIRGs is based on a luminosity criterion this may induce some apparent inconsistencies in their classification (for instance, one object could have a luminosity higher than 10$^{12}$L$_{\odot}$ [ULIRGs] for us and not for Sanders et al.). However, none of the objects in the sample were so close to the criterion limit to be in such a situation.

The morphology class was derived using mainly ground based images (Digital Sky Survey available on the NED).  For several objects these images were not enough to probe the morphologies, and additional HST images from programs 6346 (WFPC2) and 10592 (ACS, PI: A. Evans) were used.   A simplified morphological/merging classification scheme similar to that given by Veilleux et al. (2002) for ULIRGs has been used. Objects that appear to be just single isolated objects without evidence for a merging process are classified as class 0. Galaxy mergers in a pre-coalescence phase, with two well differentiated nuclei, are classified as class 1, and objects with peculiar morphologies suggesting a post-coalescence merging phase are classified as class 2 (see Fig.1 and Table 1). This classification is based on available optical imaging, and may be revisited as our program progresses. In particular, objects classified initially as 0 may show characteristics of  2 -like objects after a more detailed analysis. Therefore, since at the time of writing some objects in the sample are pending observations and reductions, it should be considered as indicative only.  

The VIMOS sample is not complete either in flux, luminosity or distance. Complete samples of (U)LIGRs, if including the whole luminosity range and morphological types,  are so numerous that they would require a prohibitive amount of observing time. (Moreover, current  reduction and analysis tools are not able to handle the large volume of data generated by very large samples of IFS data.) This is illustrated in Figure 1, which shows the need to move to higher redshifts to include higher luminosity  and morphologically diverse objects.   As mentioned above, in addition to the VIMOS sample, two additional programs in the northern hemisphere using IFS instruments (INTEGRAL-WHT and PMAS-CAHA) are being carried out by us. With INTEGRAL and PMAS we typically observe the spectral ranges 4500-7500 \AA\   and 3700-7000 \AA, respectively, with a spectral resolution of about  6\AA\   in both cases. The spatial sampling of these data  (about 1 arcsec) is somewhat poorer than the present VIMOS data.  The PMAS sample contains all the northern hemisphere targets of the complete volume-limited subsample of nearby LIRGs selected from the RBGS (Alonso-Herrero et al. 2006). The INTEGRAL sample is a representative set of 22 low-$z$ cold and warm ULIRGs covering different morphological and activity classes (Garc\'{\i}a-Mar\'{\i}n 2007). 

In terms of luminosity, the global sample of galaxies therefore contains 
three different subsamples, (i) the complete volume-limited local-LIRGs of Alonso-Herrero et al. (2006) (L-LIRG sample), (2i) the extended LIRGs (E-LIRG sample), and (3i) the ULIRG sample. The distribution of the different samples in the  L$_{ir} -z$ and L$_{ir}$ - morphology class planes are given in Figure 1, where different symbols indicate the instrument used. 

\section{Observations, data reduction, and line fitting}

\subsection{Observations}

The observations were  
obtained in service mode during semesters P76 and P78  using
 the integral field unit of VIMOS (LeF$\grave{e}$vre et al. 2003)
available at the unit 3 (UT3/Melipal) of the Very Large Telescope (VLT).  The
high resolution mode "HR Orange" (grating GG435), covering
approximately the spectral range between 5250-7400 \AA{} with a
spectral resolution of about 3400 (dispersion of 0.6 \AA~pix$^{-1}$) was used. 
In this mode the
field of view covers 27$'' \times$27$''$ with a magnification set to
0\farcs67 per fiber. For this instrument the fibers are closely packed
at the pseudoslit (mean distance between spectra at the detector is 5
pixels, with a mean FWHM of 4 pixels). This allows us to use 40 $\times$
40 fibers obtaining a total of $\sim$ 1600 simultaneous spectra per pointing.  A 
square 4 pointing dither pattern with a relative off-set of
2\farcs7, equivalent to 4 spaxels, was used per galaxy. The exposure time was set to 750
(720) s per pointing for a total integration of 3000 (2880) seconds
per galaxy during cycle 76 (78).  Weather conditions were in general
good, but vary from case to case. The mean seeing and air mass for the two 
systems analysed in
this paper were 1\farcs2, and 1.0
for IRAS F06076$-$2139  and 1\farcs5 and 1.2 for IRAS F12115$-$4656, respectively.

\subsection{Data reduction}

The VIMOS data were reduced with a combination of the pipeline
provided by ESO (version 2.0.2) via "esorex, version 3.5.1" \footnote{See
\texttt{http://www.eso.org/projects/dfs/dfs-shared/web/vlt/vlt-instrument-pipelines.html}}) 
and a set of costumized IRAF and IDL scripts. Initially the pipeline was used to perform the basic reduction steps to the individual quadrants (bias, spectra tracing and
extraction, correction of fiber and pixel transmission, and relative
flux calibration). Once the individual quadrants were partially
reduced they were combined into a single data-cube. In order to evaluate
the data reduction quality, a series of tests were performed after pipeline processing
of each individual pointing, and after generating the combined data-cube  with the four
independent dithered pointings. Results were similar for the different pointings/objects and are presented in Fig. 2 for one
of the pointings to illustrate the final quality of the calibrated data. Flux
calibration was checked by comparing the response function obtained
with the ESO pipeline and independently with the IRAF tasks \texttt{standard},
\texttt{sensfunc} and \texttt{calibrate} for one of our
objects. Differences were found to be smaller than 5\% between the two. The
\textsc{[O\,i]} 6300.3~\AA{} sky line was used to check the fiber-to-fiber
transmission correction and wavelength calibration as well as to
obtain the instrumental profile (i.e FWHM) for each spaxel/spectrum (Fig. 2   
illustrates the results obtained).
This line is the optimal one for these goals since
it is bright, isolated, and lies between the absorption NaI 5890,5896 \AA~ doublet and the
H$\alpha-$[NII] emission lines complex, which are central to the analysis. 
The effective spectral resolution was established by measuring the FWHM of a single Gaussian profile fitted to this line in each spectrum ($\sim$ 1600 in total). The spectral resolution turned out to be uniform over the entire field-of-view with an average value of 1.83 $\pm$ 0.25 $\AA$ (FWHM),
equivalent to 87  $\pm$ 12 km s$^{-1}$ at the corresponding wavelength.
The obtained mean central wavelength (6300.28 $\pm$ 0.11 $\AA$) coincides well with its actual value (6300.304 $\AA$, Osterbrock et al. 1996) 
 indicating negligible systematic errors in velocity ($\pm$ 5 km s$^{-1}$ at one sigma level). 
 Residuals in flat-fielding affecting the fiber-to-fiber flux calibration were derived measuring
 the relative flux (in counts) of the sky line for each of the spaxels in a given sky frame normalized to the median flux value of all the spaxels.
 One-sigma differences close to 30\% were found, indicating therefore that the relative flux of the same emission/absorption
  line (e.g. H$\alpha$) coming from different parts of  the same galaxy could have uncertainties of up to 30\% due to residual calibration effects. The issue of
  flat-fielding and its impact in the relative fiber-to-fiber calibration deserves further investigation for the entire set of VIMOS data.
  
Similar tests to assess the uncertainty in the wavelength calibration in the blue part of the spectral range
were performed with the 5577~\AA{} sky line. The behaviour of the wavelength calibration 
in the blue end of the spectral range is poor in about half of the cases, with errors of up to
$\sim$2~\AA. This is due to the lack of bright lines in this spectral
range for the available arc calibration lamps used during standard operations. 
For the IFU-HR\_orange mode the He 5015.675$\AA$ line is outside the covered spectral range. 
The next blue line (Ne 5400.56$\AA$) is very faint and is not identified in most of
the cases, being the next usable line at 5862~\AA. Thus, for this VIMOS
configuration, the wavelength calibration is well constrained only in the 
$\sim 5800 $-$ 7400$~\AA~ range.

Bad spaxels were identified by looking at the output tables of the
task \texttt{vmifucalib} from the ESO pipeline as well as at the
extracted continuum lamp spectra. The flat spectra corresponding to fibers
with very low transmission (smaller than 0.25 times the median
transmission) were rejected. The four pointings were combined
and the cosmic rays rejected. Because of the
dithering pattern chosen, most of the information in the dead spaxels was recovered during this
process.

Finally the sky was subtracted by obtaining a high S/N ratio sky spectrum
combining a set of spaxels not affected by the galaxy emission. In those few
cases where the galaxy covered the whole VIMOS field of view,
the sky spectrum obtained for a galaxy observed during the same night was used,
scaling its flux until the sky line residuals in the subtracted image
in the red part of the spectrum were negligible.
The final "supercube" has 44$\times$44 spaxels (1936 spectra) .

 \begin{figure*}
  \centering
  \includegraphics[angle=90,width=18cm]{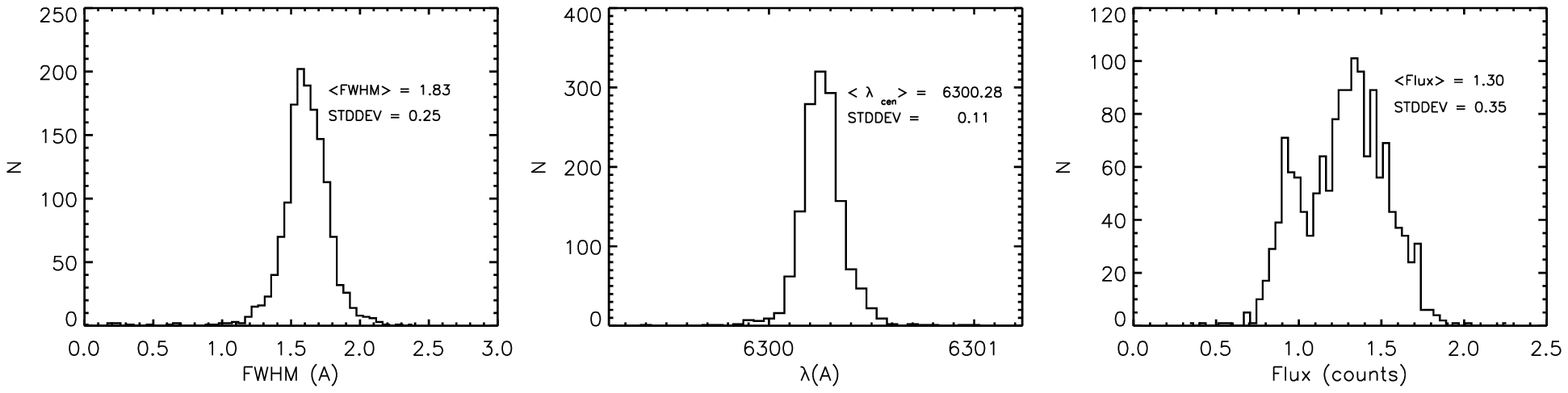}
     \caption{Histograms showing the results of the calibration checks using 
     the [OI] sky line  at 6300.3 $\AA$
for the first pointing of
IRAS F06076$-$2139, after processing the data through the
pipeline. \emph{Left:} line (instrumental) width distribution for all the spaxels, and 
binned in 0.05 \AA-width intervals. \emph{Middle:} Distribution of the
wavelength for the centroid of the sky line. \emph{Right:} Distribution of the
total line flux normalized to the median flux of the spaxels.}
        \label{FigVibStab}
  \end{figure*}

\subsection{Line Fitting}

The 1936 spectra of each "super-cube" data were inspected visually in
order to check for possible reduction artifacts, as well as for the presence
of multiple components, peculiar line profiles, etc. In order to extract the
relevant emission line information, line profiles were fitted with a
Gaussian model. A single Gaussian profile per emission line was
sufficient in most of the cases, although occasionally  two, or even three
components per line were needed. Therefore, in a first step, we fit
automatically all the lines to a single Gaussian. Then, if necessary,
multi-component fits were performed in some spaxels restricted to 
certain regions of the field-of-view. These steps were carried out with the DIPSO
package (Howarth \& Murray 1988) and the MPFITEXPR algorithm implemented by C.B. Markwardt in the IDL environment\footnote {available at http://cow.physics.wisc.edu/$\sim$craigm/idl/idl.html} , which allows us to fix wavelength
differences and line intensity ratios according to atomic parameters
when adjusting multiple emission lines such as the
H$\alpha$-\textsc{[N\,ii]} spectral range. Several tools have been developed 
to perform these tasks as automatically as possible. By default the
same line width was considered for all the lines.  For each
emission line we ended with the following information: central
wavelength, FWHM, and flux intensity. The central wavelengths were
translated into heliocentric velocities taking into account the radial
velocities induced by the Earth motion at the time of the
observation. These were corrected with the IRAF tool
\texttt{rvcorrect}. The FWHM were corrected spaxel-to-spaxel by the 
corresponding instrumental profile (average of 1.83 $\pm$ 0.25 \AA, see Fig. 2)
obtained from the [OI] 6300$\AA$ sky line.

\section{First results and discussion}

In this section we present the first VIMOS results obtained for this program. Here we illustrate the full potential of such a survey by analyzing two representative galaxies. The whole sample will be analysed in a series of future publications. The galaxies presented here were selected taking into account their different morphological and kinematic characteristics. On the one hand, IRAS F06076-2139 is a system with a rather complex morphology showing the presence of two clearly distinct galaxies that appear to be interacting.  On the other hand, IRAS F12115$-$4656 is a spiral galaxy which is in (weak) interaction with a relatively distant companion located at about 100 kpc (projected) away. Their infrared luminosities are   10$^{11.67}$ and 10$^{11.11}$ L$_{\odot}$, respectively, and therefore they span a factor of 3.6 in luminosity within the range covered by LIRGs. They may be considered representative of LIRG sub-types  and, as we will show in the following, the VIMOS data have revealed interesting new characteristics in both cases. The two systems are discussed separately in the rest of the paper.

\subsection{IRAS F06076$-$2139} 

The ground-based optical (Digital Sky Survey)  and near-infrared (2MASS) images obtained from the NASA Extragalactic Database (NED) for this object show a relatively simple morphology: a double-nucleus embedded in an elliptical-like galaxy.  However, the recently avaliable archival ACS/HST images (program 10592, PI: A. Evans) reveal a very  rich and complex morphology of two clearly distinct galaxies  (hereafter referred to as G$_n$ and G$_s$ for the northern and southern galaxy, respectively) with a projected nuclear separation of about 9 arcsec (i.e. 6.7 kpc). The ACS I-band image (i.e. filter F814W) is presented in Figure 3 as a reference. The galaxy G$_n$ shows a diffuse double ring (ribbon-like), and a spiral of bright distinct knots in the innermost regions, crossed by a prominent dust lane. On the other hand, G$_s$  has a very prominent inner ring of knots of about 0.5$''$ (370 pc) in radius (hereafter referred to as the $\it {crown}$, to be distinguished from an outer ring -- see below) and a more extended envelope, elliptical in shape.  
Of particular interest in the morphology of galaxy G$_s$ is the presence, as inferred from the ACS images, of an external ring structure with a semimajor axis of 6 arcsec
(i.e. 4.5 kpc), concentric to the nucleus, and with several blue compact knots at different angular locations.  Finally, many faint blue knots are 
detected around the main body of the system, in particular towards the north and northwest
(see Fig. 3).

\subsubsection {Spatial structure of the emission line profiles. Evidence of the kinematic and ionizing complexity}

The spectra show prominent changes in the line profiles, velocity shifts and emission line ratios, from region to region within the system (see Figure 3 for some examples of selected regions). 

   \begin{figure*}
   \centering
      \caption{{\bf  FULL VERSION at  http://damir.iem.csic.es/~aalonso/extragalactic/publications/publications.html} VIMOS spectra in the redshifted H$\alpha$-[NII] spectral range from selected regions of IRAS
       F06076-2139, as indicated in the HST-ACS F814W-band image, which is represented on a logarithmic scale 
      to enhance the low intensity regions. (This image is a subsection of the calibrated image as provided by the HST archive, after rotating and removing the cosmic rays using standard IRAF tasks.) The maximum (white) and minimum (black) correspond to 6 and 0.1 in arbitrary flux units. Numbers inside the panels identify the spaxel. The spectra shown illustrate the changes in velocity, line
       profiles and emission line ratios.
      The yellow frame indicates the VIMOS  field of view represented in other figures of this galaxy  in the paper (the actual
       VIMOS field of view is somewhat larger). The small square in the right lower corner of the figure shows
        the size of a VIMOS spaxel  (0.67$''$ $\times$ 0.67$''$). The arrow indicating the orientation has a
         length of 5$''$. The  symbol RK identifies the red knots (see text). }
         \label{Esp.}
   \end{figure*}

The spectra of the two nuclei  associated with the two galaxies of the system (N$_n$ and N$_s$ in Figure 3, respectively) show emission lines with a large relative velocity difference of $\sim$  550 km s$^{-1}$, and with different line profiles.  While N$_n$ shows relatively broad lines ($\sigma$=104 km s$^{-1}$) with a low intensity blue wing, the southern nucleus (N$_s$) has  narrower lines ($\sigma$=63 km s$^{-1}$). There are some other regions (spectra [18,24] and [8,23] in Figure 3) where the presence of secondary blue components (i.e. blueshifted systems) is clear. Since these regions are spatially separated by large distances (about 5 kpc), and show  large velocity differences, these secondary kinematic components could be associated with  weak emission from the other galaxy of the system seen in projection along the line of sight. In fact, in certain spaxels (spectra [17,26]) two main kinematic systems are identified, one per galaxy.   Interestingly, the [NII]/H$\alpha$ ratio is quite different for these two main velocity systems (see differences between the [18,24] and [8,23] spectra). Finally, there are some other spectra in the outer regions of the system with narrow lines associated with either diffuse gas within the external ring of G$_s$ ([7,35]), the compact star-forming knot north of  G$_n$ (spectrum [30,24]), diffuse ionized gas in the outer regions of the system (14,5] and [13,7]), or bright blue knots ([10,5]) . Two knots denoted by "RK" outside the VIMOS field of view show very red colours (B-I $\sim$ 3.5-4.1), as inferred from the F435W and F814W ACS images.   

\subsubsection {Stellar and ionized gas distributions. Detecting new structures in the system}

In order to fix the coordinate system for the VIMOS spectral maps into the HST reference system, a
VIMOS emission line-free, stellar continuum image (Figure 4, top-left) was generated by integrating the spectra in the observed 6629 - 6733 $\AA$ spectral range (rest frame 6390 - 6490 $\AA$).  Aside from the steps mentioned in Section 3, some spaxels of this image had to be cleaned similarly to the standard cosmic ray rejection in standard imaging.  This map follows the global structure of the ACS image although, obviously,  with poorer spatial resolution. However, the patchy structure around the nucleus of the northern galaxy and the presence of the crown in the southern nucleus introduces some uncertainties in the VIMOS- HST reference systems, which have been estimated to be $\pm$ 0.3$''$. 
The other panels shown in at the top row of Figure 4 were obtained fitting a single component per emission line. They correspond  to H$\alpha$, [NII]/H$\alpha$ line intensity ratio, mean radial velocity (v), and velocity dispersion ($\sigma$).

   \begin{figure*}
   \centering
      \caption{{\bf  FULL VERSION at  http://damir.iem.csic.es/~aalonso/extragalactic/publications/publications.html} VIMOS results for IRAS F06076$-$2139. Upper panels represent the stellar continuum (left) and H$\alpha$ light (second left) distributions. Changes in the 
      ionization conditions are traced by the  [NII]$\lambda$6584/H$\alpha$ line ratio (central). The velocity field (second right) and velocity dispersion map (right) are also presented. The central
      and lower rows represent the same characteristics for the two components identified in certain regions of the galaxy through a multi-component emission line fit to the observed
      spectra (see text for details). The continuum and H$\alpha$ maps are represented on a logarithmic scale between 0.63 (red) and 0.02(black), and 40 (red) and 0.03 (black), respectively,  in arbitraty flux units. }
         \label{FigVibStab}
   \end{figure*}

The structure of the ionized gas traced by the H$\alpha$ line looks very different from the stellar light
distribution mentioned above. Galaxy G$_n$ shows a more compact H$\alpha$ morphology than the stars with the H$\alpha$ emission centered on and dominated by the nucleus of the galaxy. On the other hand, G$_s$ exhibits not only emission associated with the nucleus (most likely associated with the "{\it crown}" observed in the ACS image) but also a prominent ring of ionized gas of about 16 arcsec in diameter (i.e. 12 kpc) and with its major axis along PA $\sim$ 13 degrees. In addition, a chain of several H$\alpha$ clumps forming a bended discontinuous (broken) tidal tail-like structure is detected at about 14-16 arcsec ($\sim$10.5 $-$ 12. kpc) west and northwest of the nucleus of  G$_s$. All of these H$\alpha$ clumps, which are not detected in the continuum map,
show a remarkable positional coincidence with the external blue faint knots detected in the
ACS F435W image (see Figure 5), indicating a clear physical association between them. Therefore, these H$\alpha$ clumps must be
young star-forming regions, or tidal dwarf galaxy (TDG) candidates, formed as a consequence of a collisional encounter.

  \begin{figure*}
   \centering
      \caption{{\bf  FULL VERSION at  http://damir.iem.csic.es/~aalonso/extragalactic/publications/publications.html} H$\alpha$ emission (contours) as derived from the VIMOS-IFU data on top of the ACS/HST blue band (filter F435W)  image 
      showing the remarkable positional coincidence of the external H$\alpha$ chain of clumps, detected with VIMOS, with the faint, blue stellar blobs located outside the main system formed by G$_n$ and G$_s$.   The contours  span from  25 to 0.03 in arbitrary flux units in a logarithmic scale. The colors  range from 10 (white)  to 0.03 (black) in arbitrary flux units in a logarithmic scale.}
         \label{FigVibStab}
   \end{figure*}

The observed H$\alpha$ flux is dominated by the two nuclei of the galaxies which contribute 39\% (northern nucleus) and 29\% (southern nucleus) of the total measured flux, respectively. The ring accounts for 22.5\% of the H$\alpha$ flux while the rest (9.5\%) comes from the external chain of H$\alpha$ clumps. Without a correction for internal extinction, the total observed H$\alpha$ luminosity in the system is equivalent to a star formation rate of 1 M$_{\odot}$
yr$^{-1}$, almost two orders of magnitude lower than the value of 81 M$_{\odot}$ yr$^{-1}$ derived from the
infrared luminosity (see Table 2 for details). Thus, since extinction in (U)LIRGs tends to increase inwards, it is likely that the H$\alpha$ emission is being absorbed preferentially in the two nuclei. From the ACS images (see Figures 3 and 5), the presence of dust lanes is very prominent in the (circum)nuclear regions of the two galaxies. Therefore the observed H$\alpha$-derived SFR value would reach an agreement with the IR-derived value if a visual extinction of about 5 magnitudes exists in at least one of the nuclei of the galaxies. Visual
extinctions measured in the central regions of LIRGs are in the 2 to 6 magnitudes range (Alonso-Herrero et al. 2006).
 
On the other hand, the total H$\alpha$ luminosity measured in the external H$\alpha$ clumps corresponds to 1.3 $\times$ 10$^{40}$ erg s$^{-1}$, with individual clumps (at the VIMOS angular resolution) having H$\alpha$ luminosities in the 0.2 to 0.6  $\times$ 10$^{40}$ erg s$^{-1}$ range. In addition, each of the blue-continuum knots associated with these clumps have an effective radius of about 100 to 200 pc (measured directly from the ACS B-band image).  These properties are similar to those measured in some of the TDG candidates detected in ULIRGs such as IRAS F08572+3915 and IRAS F15250+3609  at higher redshifts (Monreal-Ibero et al. 2007).  

Finally, the ionization structure of the gas traced by the [NII]$\lambda$6584/H$\alpha$ line ratio (Figure 4, top-center panel) is clearly distinct between the
northern and southern galaxies. The southern nucleus together with the H$\alpha$ ring and external clumps all have an HII-like excitation. However, the nucleus and circumnuclear regions of the northern galaxy have ratios changing (in log scale)  from $-$0.12 (nucleus) to extreme values of +0.8 at about 4 arcsec (3 kpc) from it. Additional line ratios measured in this region  (log([SII]/H$\alpha$) = +0.60) are also high, well above the limits of AGN-like galaxies defined by a sample of 85224 low-redshift emission-line galaxies from the Sloan Digital Sky Survey (Kewley et al. 2006). The average spectra in this region show a weak H$\alpha$ in emission on top of a stellar H$\alpha$ absorption line. When corrected for this absorption, the [NII]/H$\alpha$ and [SII]/H$\alpha$ ratios (in log scale) move to +0.4 and +0.2, respectively. These ratios correspond to high excitation Seyfert (i.e. AGN), LINER-like (weak-AGN or shocks) regions, or a composite type of ionization. Although a measurement of the [OIII] 5007\AA~ line is needed to distinguish between these two mechanisms, the off-nuclear location of these zones could suggest that shocks, as in many other (U)LIRGs (Monreal-Ibero et al. 2006; Garc\'\i a-Mar\'{\i}n 2007), are playing a dominant role in these regions.

\subsubsection {Ionized gas kinematics: Towards understanding the nature of the system}

The velocity map (Figure 4, second to the right top panel) shows several interesting characteristics that give hints about the nature of this system. 
First, the measured H$\alpha$ velocity of the nuclei of the galaxies (V(Nn)=11186 $\pm$ 25 km s$^{-1}$ and V(Ns)=11737 $\pm$ 25 km s$^{-1}$) clearly reveals 
the presence of two galaxies moving with respect to each other with a velocity of about 550 km s$^{-1}$.  Since this is a lower limit to their actual relative velocity,  it is unlikely that these two galaxies form a real merging pair. 

Second, the northern galaxy G$_n$ has an internal velocity gradient (peak-to-peak) of about 150 km s$^{-1}$ over 6 arcsec (4.5 kpc) but with the kinematic major axis tilted by about 60 degrees with respect to the photometric axis. 
The fact that this velocity field is consistent with that of a massive, rotating disk of about 5 kpc in size
and tilted with respect to the stellar mass distribution (see $\S$ 4.1.6) would indicate an external origin of the gas.

Third, the southern galaxy G$_s$ has a regular velocity field with the only pecularity that while the kinematic major axis of the (circum)nuclear regions (inner 4$"$) is oriented along a PA of  $\sim$ 0 degrees, i.e.
close to the stellar major axis of the galaxy,  the major kinematic axis of the ring is however oriented nearly perpendicular to its photometric major axis. This suggests that IRAS F06076--2139 is a collisional ring galaxy (e.g. Elmegreen \& Elmegreen 2006, and references therein). 

Finally, the external H$\alpha$ clumps at distances of
about 12 kpc from the nucleus of G$_s$ shows a rather constant velocity, independent of their orientation and in agreement with the velocities measured in the northwestern side of the H$\alpha$ ring. Thus, this large scale velocity structure suggests that the southern galaxy, the ring and the external clumps  are part of the same galaxy that has been involved in an interaction, most likely head-on as indicated by the velocity structure of the ring (see $\S$4.1.5)

The velocity dispersion ($\sigma$) map (Figure 4, top-right panel) also shows a clear distinction between the two galaxies, with G$_n$ exhibiting consistently larger dispersions. This suggests that this galaxy is more massive than G$_s$, as also indicated by the light distribution.  

Although the northern nucleus (N$_n$) is located in a local maximum of 104 km/s (i.e. it has broader lines than its inner circumnuclear region), it is not well centered on the absolute maximum.  Actually, the broadest lines are found in the  transition region between the two galaxies, where two components of similar flux but shifted in velocity are present. The velocity dispersion of the nucleus of  G$_s$ (N$_s$) has a value of  63 km/s, while in the ring and chain of clumps is somewhat lower ( 35-40 km/s), indicating a more relaxed, quiescent gas.   The very large $\sigma$ values measured in a region located  $\sim$ 3 arcsec SW from the nucleus of G$_s$ are due to the the presence of many faint blueshifted components of gas, likely connected to the G$_n$. These components are probably present along  other lines of sight, but they are only noticeable when the component of the southern nucleus is faint enough.

\subsubsection {Kinematic deprojection. Origin of the multiple ionized gas velocity components} 

As mentioned above, IRAS F06076-2139 shows in some regions multiple components in the emission lines. Although a detailed analysis of the multiple kinematically distinct components can be extremely complex (and may require higher spectral resolution), the fact that two main components (in some cases with clear evidence of subcomponents) dominate the emission simplifies this study. Therefore, in the cases where only one single Gaussian per emission line was not a good representation of the observed line profile, two Gaussians per line were fitted instead. Note that the spatial identification of the 
components ({\it kinematic deprojection}, see Arribas et al. 1996) can be done unambiguously thanks to the fact that they have quite different kinematic properties (i.e. mean radial velocity and velocity dispersion). Therefore, in general, at each spatial position there was no doubt as to which fit should be associated with which component.  This allows us to construct spectral maps of each kinematically distinct component.  The central and lower rows in Figure 4 represent the spatial distribution of H$\alpha$, the [NII]/H$\alpha$ line intensity ratio, mean radial velocity (v), and velocity dispersion ($\sigma$) for the two components. The double components are found only in two regions: i) the (projected) interface region between both galaxies, and ii) westwards of the southern nucleus.
A third region with double components was also found to the north of G$_n$ but it has not been considered here due to the large uncertainties in the fits with two components. 

Looking at these spatial distributions the so called components 1 and 2  share the kinematic and ionization properties of galaxies G$_n$ and
G$_s$, respectively, and are therefore physically associated with each of them. In fact component 2 (Figure 4, bottom panels) has the line widths, mean radial velocities, and [NII]/H$\alpha$ ratios expected for G$_s$ at the spatial location where it is found. Similarly, the properties of component 1 (Figure 4, central panels) are those expected for G$_n$.   However, interestingly, the radial velocities of the region close to the southern nucleus are intermediate between those expected for the two galaxies.

\subsubsection {Kinematic evidence for an expanding collisional ring in G$_s$}

One of the most interesting kinematic features of the velocity field associated with G$_s$ is the regular pattern of the radial velocities in the H$\alpha$ ring. The fact that the largest velocity gradient is nearly perpendicular to the photometric major axis of the ring suggests that the ring may have a peculiar velocity pattern with an important expansion component. 

As a first step we selected the spaxels/spectra containing information about the ring. For those spectra showing two components per emission line, only the one corresponding to the ring was considered (see previous subsection).   To analyze in detail the velocity field of the ring, a simple kinematic model (Mihalas \& Binney, 1981) was fitted to the data:

 $v_r (R, \theta$)= [$\Pi (R,\theta) \times sin (\theta-\theta_o) + \Omega(R,\theta) \times cos (\theta-\theta_o)] \times sin i + Z (R,\theta) \times cos i + V_{sys}$,                               (1)

\noindent
where $\Pi(R,\theta)$, $\Omega(R,\theta)$, and $Z(R,\theta)$ are radial, azimuthal, and perpendicular components of the velocity field, $i$ is the inclination angle (defined as the angle between the normal to the galaxy disk [ring] and the line of sight),  $V_{sys}$ is the systemic velocity, $\theta_o$ the position angle of the projected major axis of the galaxy (ring), $R$ the galactocentric distance , and $\theta$ the galactic azimuthal angle.  This model is rather generic, but it can be simplified for the present case taking into account some observational facts. First,  our data indicate azimuthal symmetry and therefore the amplitude of the  of the velocity components does not depend on the galactic azimuthal angle. Second, the global systemic velocity of the ring is in agreement with the radial velocity of the nuclear region (see Table 2), indicating that the value of the perpendicular component (i.e. $ Z $) is negligible. Taking into account these facts, equation (1) can be reduced to: 

 $v_r (\theta)= [\Pi (R) sin (\theta-\theta_o) + \Omega(R) cos (\theta-\theta_o)] \times sin i  + V_{sys}$,       (2)

 \noindent
where $\theta_o$ can be easily inferred assuming that the actual geometry is circular.   The observed position angle inferred from these images is 13$^o \pm 2$. The measured values for the lengths of the major and minor axes of the observed ring are then used to derive the angle of inclination, $i$, which is found to be 57$^o \pm 2$ (with a global uncertainty of 180$^o$). 

\begin{figure}
    \centering
   \includegraphics[angle=-0,width=10.5cm]{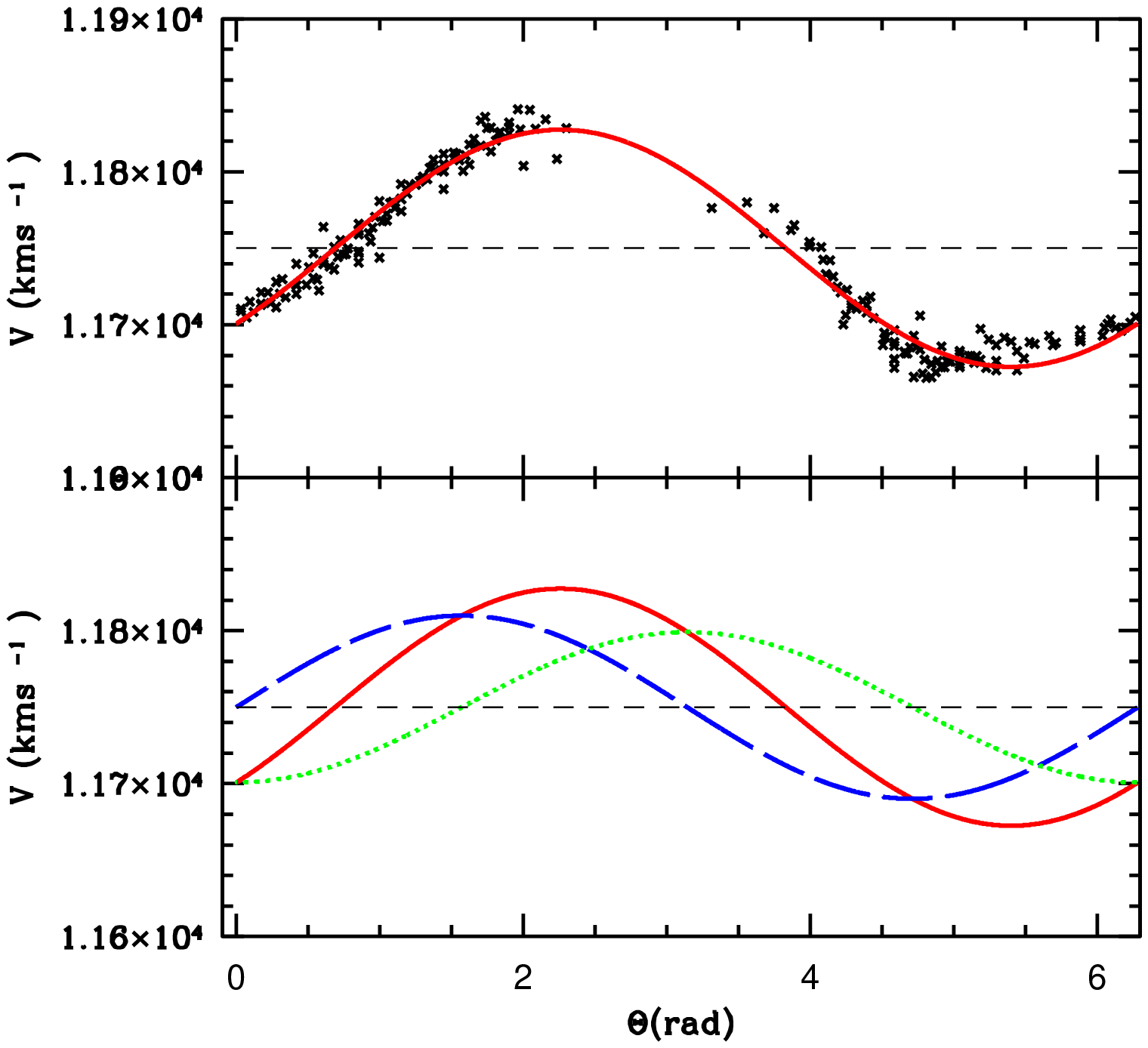}
      \caption{Top. Best fit  (red line) and data points for the kinematic model of the ring in G$_s$ as  described in
       equation (2).  Bottom. Decomposition of the total radial velocity of the modeled ring (red line) into its
        orbital (green) and azimuthal (blue) components. The dashed line shows the systemic velocity of the
         ring.                 }
         \label{FigVibStab}
   \end{figure}
   
Considering these values, equation (2) is solved for $\Pi$, $\Omega$,  and $V_{sys}$ by applying a least square fit to the data (see Figure 6, top panel, for the 206 data points and the best fit). The best fit (Figure 6, bottom panel) was found for 
an expanding velocity $\Pi$ = 70 $\pm$ 3 $kms^{-1}$, a clockwise rotation $\Omega$ = 61$ \pm$ 3 $kms^{-1}$,  and a systemic velocity $V_{sys}$ = 11750 $\pm$ 1 $kms^{-1}$.  For  $\Pi$  and $\Omega$ the errors were estimated from the uncertainties  associated with the $\theta_o$, and $i$ only. 
 For $V_{sys}$ it was estimated from the standard deviation of the differences data - model, which was 0.7 $kms^{-1}$. All these errors do not include systematic effects due to both the measured data, and the model approach. Mathematically the radial component,  $\Pi$, could be explained by both expansion or contraction (due to 180$^o$ uncertainty in $i$). However, from a physical point of view, we can safely exclude radial contraction at such a velocity. Furthermore, in ring galaxies  we actually expect the presence of expansion induced by the  wave created in reaction to the impact. The fact that the ring is expanding fixes its own geometry in such a way that  its SW sector must be the one closer to us. That in turn fixes the direction of the azimuthal (rotational) component which is clockwise. This result, which relies on solid kinematical arguments, seems to contradict the morphological  spiral appearance of some structures inside the ring in the ACS B-band image, which suggest an anti-clockwise rotation  if it is assumed that these structures are trailing. However, this apparent contradiction can be explained as differential rotation inside the ring producing a rising 'rotation curve' such that the outer parts are rotating faster than expected for solid-body rotation.

\subsubsection {Evidence for a gaseous kinematically decoupled rotating disk in G$_n$}
   
The velocity field of G$_n$ does not show a clear large-scale rotational pattern.  The kinematic major axis seems to be tilted with respect to the stellar photometric axis, and the velocity dispersion map is also difficult to interpret as due to pure rotation. The ACS images also shows a quite complex stellar light distribution with several dust lanes and what appears to be star-forming regions in a spiral-like structure (Figure 3).  We attempt  to fit  the velocity field to the model described above (eq.[3]). As a first order approximation, we considered a rotational model with a constant value for  $\Omega$.  The best fit model for $i$=58 degrees (derived from the ACS images) and considering $\Omega, \theta_o$, and  $V_{sys}$ as free parameters  gave a rotational velocity of 76 km s$^{-1}$, a position angle for the major axis of +40 degrees, and a systemic velocity of 11236 with an rms of $\sim$ 3 km s$^{-1}$.  If  $\theta_o$ is fixed to  $-$20 degrees (i.e. the stellar photometric axis position as derived from the ACS images) the fit has residuals substantially larger than the model above.  These results indicate that, despite the relatively large deviations, the velocity field in galaxy G$_n$ is consistent with rotation around an axis tilted by about 60 degrees from the stellar major axis. This suggests that the ionized gas is decoupled from the stars. Moreover,  the high velocity dispersions outside the nuclear regions also indicate that the ionized gas in G$_n$ has quite complex kinematics.  The distribution and kinematics of the ionized gas and the stars are often known to be decoupled in early-type galaxies (e.g., Bertola \& Corsini 1999).  The velocity fields show misalignments between the stars and the gas (e.g., Kannappan \& Fabricant 2001;   Sarzi et al. 2006). This has been used to constrain the origin of the gas in early type galaxies. In the case of IRAS F06076$-$2139, which clearly shows strong merging/interaction features, it is not surprising to find such misaligment.

\subsubsection {IRAS F06076$-$2139. Overall kinematic model and physical parameters}

 \begin{figure*}
    \centering
      \caption{{\bf  FULL VERSION at  http://damir.iem.csic.es/~aalonso/extragalactic/publications/publications.html} Comparison of the H$\alpha$ observed (left) and modeled (centre) velocity field for IRAS F06076$-$2139. The modelled velocity field  combines the model of the tilted rotating disk (case 1, see text) in G$_n$ with the expanding, rotating ring in G$_s$. The  residual (observed $-$ modeled) velocity field is also given (right). The central spaxel at (0,0) corresponds to the nucleus of  G$_n$, and has been excluded since it is a singularity for the model.}
         \label{FigVibStab}
   \end{figure*}

 The expanding rotational ring model for the southern galaxy has been combined with the tilted rotational disk for the northern galaxy in order to compare the overall model with the observed velocity field of the system  (see Figure 7). 
 The model does not include the internal regions of the southern galaxy due to
 the limited angular resolution and to the fact that the ACS images show the presence of a nuclear star-forming ring (the crown) that could have its own velocity field independent of the one detected in the outer expanding ring.
 The model explains well the overall velocity field on large scales as the average residual (observed $-$ model) is $-$4 $\pm$ 11 km s$^{-1}$ and +0.6 $\pm$
 29 km s$^{-1}$ for G$_s$ (ring) and G$_n$, respectively. However, significant residuals, blueshifts larger than  $-$ 50 km s$^{-1}$ with respect to rotation, appear  on smaller scales in G$_n$.
  The nature of these shifts is unclear, but certainly demonstrate that the velocity field in G$_n$ is not only due to the pure rotation of a tilted disk but is also affected by additional radial flows, or tidally-induced motions.

From the azimuthal velocity component (i.e. rotation), $\Omega$, we can estimate the mass inner to the ring (i.e. 10.1 kpc) in G$_s$, at a radius of 5.6 kpc for G$_n$. The dynamical masses inside these radii  are M(G$_s$)= 1.7 $\times$ 10$^{10} M\odot$ and M(G$_n$)= 7.5 $\times$ 10$^9 M\odot$, respectively. However, these masses represent however only a fraction of the overall mass of the system. Under virialization and considering a polytropic mass distribution, the overall mass of each of the galaxies is given as 1.24 $\times$ 10$^{-5}$ $\times$ R$_{hm}$(kpc) $\times$ $\sigma^2$(km s$^{-1}$) (see Colina et al. 2005 for details) in units of an m$_*$ galaxy
(1.4 $\times$ 10$^{11} M\odot$, Cole et al. 2001).  Assuming that light traces mass, the half-mass radius can be obtained measuring the effective radius. For this purpose, polygonal apertures equivalent to circular apertures of 12.8$"$ and 8.6$"$ in radius for G$_n$ and G$_s$ respectively, have been used in the recently available ACS/F814W image. The polygonal apertures allow us to better estimate the effective radius of each of the galaxies as the aperture is selected in such a way as to minimize the light contribution from the companion. The corresponding effective radii are 2.76 kpc and 2.73 kpc for G$_n$
and G$_s$, respectively. Taking into account the measured central velocity dispersions of 
104 km s$^{-1}$ and 63 km s$^{-1}$,  IRAS F06076$-$2139 is formed by  two low-intermediate mass galaxies with masses of 5.2 (G$_n$) and 1.9 (G$_s$) $\times$ 10$^{10}$ M$_{\odot}$, i.e. 0.4 and 0.15 m*, respectively.

Finally, the expansion velocity of the ring can be used to date the epoch when the impact that generated the ring took place. Assuming that the ring has been expanding at a constant speed of 70 km s$^{-1}$ and that the distance to the nucleus is 10.1 kpc,  such an impact happened about  1.4 $\times$ 10$^8$ years ago.  By measuring the "width" of the ring, one can also estimate the spread in age of its  star forming regions, which is $\sim$ $2 \times 10^7$ years.

\subsubsection {Status and evolution of IRAS F06076$-$2139}

Based purely on the morphology as seen in the high angular resolution ACS images, IRAS F06076$-$2139
would be considered a merging system formed by two bright spirals. However, the VIMOS IFS provides a set of new quantitative data that challenges this simple scenario, and shows the full complexity of the system. 

On the one hand, the newly identified H$\alpha$ ring in G$_s$ has been interpreted as an expanding rotating collisional ring. Such rings are generally produced by head-on collisions with an intruder having a mass lower than the victim. Additional morphological and kinematic features of the ring give new information about this collision. 
The collision happened about 140 million years ago as derived by the size and expansion velocity of the ring.
The fact that the nucleus of G$_s$ lies close to the center of the ring indicates that the passage of the intruder was near-central. Moreover, the expansion velocity of the ring is relatively high (70 km s$^{-1}$) and about  1.15 times its rotational velocity. Since the amplitude of the expansion velocity is directly proportional to the mass of the intruder (Struck-Marcell \& Appleton 1987), the intruder should have had a mass smaller than but close to that of G$_s$. These new facts rule out a direct collision between G$_n$ and G$_s$ as the cause of the ring since: i)  the intruder (remnants) should be located along the minor axis of the ring (G$n$ is more than 70 degrees off from that axis),  and ii)  G$_n$ is considerably more massive than G$_s$ (by a factor of 2-3), according to the different tracers used.  
The brightest of the blue external knots identified in the ACS B-band image, and also detected as an H$\alpha$ emitter with VIMOS, is located at about the right distance and along the minor axis of the expanding ring, as expected for the intruder (see Figure 5). However, this knot appears to be a member of the chain of several clumps detected north and northwest of G$_s$ sharing similar kinematics and ionization properties and without evidence of an old stellar population (traced by the I-band light)  that could be considered the renmant of the intruder.  However, the red knots found by comparing the F435W and F814W ACS images (see Fig.3, labels RK), with B-I colors of 3.5 and 4.1,  are good candidates for the remnants of the intruder. Unfortunately they are outside the VIMOS field of view, and therefore we lack the kinematic information to check this hypothesis. 

As for the fate of this system, the high relative velocity  between the two galaxies is larger than the escape velocity considering  the galaxy masses inferred from the present observations. Therefore, it is very unlikely that these two galaxies will ever merge, although they could pass through each another.  Without knowing their actual distance, nor their relative velocity in the plane of the sky it is difficult to predict the future evolution with certainty.

\subsection {IRAS F12115$-$4656}

This galaxy is member of a pair included in the catalogue of interacting galaxies of Arp \& Madore (1987) with the second member of the system (IRAS F12112$-$4659) located at a distance of about four arcmin (89 kpc) SW of IRAS F12115$-$4656. The emission line spectrum of the nucleus has been classified as HII-like, and its absorption spectrum as due to a young stellar population (Sekiguchi $\&$ Wolstencroft 1992).  IRAS F12115$-$4656 was also included in the study of Palunas and Williams (2000), who presented axisymmetric maximum disk mass models for a large sample of spiral galaxies based on rotation curves extracted from their two-dimensional Fabry-Perot  H$\alpha$ data (Schommer et al. 1993). For the present paper, which presents the first results from the VIMOS sample, this galaxy was selected mainly because it is representative of a class of local LIRGs with no clear morphological indication of being involved in a large gravitational interaction or merging process. 
Unlike IRAS F0607$-$2139,  IRAS F12115$-$4656  is classified as an early type spiral (SA(rs)ab, Buta 1996).  


\subsubsection {Stellar and ionized gas distributions. Tracing the ionization sources}

The stellar light distribution was obtained generating a narrow emission line-free continuum image in the observed 
6508 - 6610 $\AA$ spectral range (Figure 8, left panel). The bright knot at about 10 arcsec NE from the nucleus is a field star. 
The emission lines in the spectra of  IRAS F12115$-$4656  show in general simple profiles with only one kinematic component per line. However, some spectra have a faint red (blue) wing predominantly in the NW (SE) of the nucleus. In a few cases we found some hints of double components, but the relative velocity between these components is at the limit of being resolved with the resolution and S/N level of the present data. Therefore, all the spectra were satisfactorily fitted with a single Gaussian per emission line. From these fits the H$\alpha$ light distribution is obtained (see Figure 8, second left panel). 

The general morphology of the ionized gas (traced by the H$\alpha$ map) follows that of the stars. However
there are some clear differences. The stellar distribution shows an elliptical shape with a well defined central nucleus which appears as a relatively weak H$\alpha$ emitter. The circumnuclear regions show, on the other hand, the presence of compact, bright H$\alpha$ knots. These knots at distances of 1 to 1.5 kpc are most likely HII regions located in the inner spiral arms of the galaxy. Also two regions at about 3.7 kpc southwest of the nucleus, and undetected in the continuum, are clearly distinct in  H$\alpha$. Finally, two faint H$\alpha$ emitting regions at  5.1 kpc NW and SE  the nucleus are also visible.  

The location and nature of the ionizing sources are traced by the [NII]/H$\alpha$ line ratio (Figure 8, central panel). According to their ionization status, three main, distinct regions can be identified: 
(a) the H$\alpha$-weak nucleus, (b) the circumnuclear HII regions, and 
(c) the elliptical outer ring of enhanced [NII] emission. 
The weak-H$\alpha$ nucleus has  a relatively high [NII]/H$\alpha$ value (log[NII]/H$\alpha$ $\sim$ $-$0.15).
This ratio together with other ionization tracers (log[OI]/H$\alpha$ of $-$1.31 and log[SII]/H$\alpha$ $\sim$
$-$0.39) locates the nucleus in the borderline region of Seyfert/LINERs and HII regions, according to the standard emission line diagrams (Kewley et al. 2006). 
Additional [OIII]/H$\beta$ line ratios would be required to establish a more precise classification.  The circumnuclear H$\alpha$ knots, and the faint two regions located SW of the nucleus have line ratios (log[NII]/H$\alpha$ of $-$0.37, log[OI]/H$\alpha$ of $-$1.67 and log[SII]/H$\alpha$ $\sim$
$-$0.55) typical of HII regions. Finally,  the elliptical outer ring clearly defines the boundary of the main body of the galaxy in the H$\alpha$ map. In this elliptical ring the average log[NII]/H$\alpha$ ratio is $-$0.02, typical of LINER or Seyfert-like ionization. The same elliptical structure is also identified in the velocity dispersion map (Fig. 8 right panel) indicating that there is a clear spatial correlation between the ionization status of the gas and its kinematics. Similar correlations have been found in the outer parts of ULIRGs, being explained by the presence of local ionizing sources (shocks) in these regions, rather than by the ionization due to a central, weak AGN (Monreal-Ibero et al. 2006).

 \begin{figure*}
    \centering
      \caption{{\bf  FULL VERSION at  http://damir.iem.csic.es/~aalonso/extragalactic/publications/publications.html} VIMOS results for IRAS F12115$-$4656 showing the stellar continuum (left), and the warm ionized (second left) light distributions together with the ionization map traced by the ratio of the [NII] to the H$\alpha$ lines (center). Also shown are the velocity field (second right) and velocity dispersion map (right) as obtained from the fit of the H$\alpha$ line to one component Gaussian profile. The continuum and H$\alpha$ maps are represented in a logarithmic scale between 12.5 (red) and 0.16 (black), and 40 (red) and 0.03 (black), respectively, in arbitraty flux units.}
         \label{FigVibStab}
   \end{figure*}

Assuming that the gas is ionized only by young stars, the equivalent star formation rate associated with the observed H$\alpha$ flux corresponds to 1.9 M$_{\odot}$ yr$^{-1}$, or
about 10\% of the total SFR derived from the IR luminosity (see Table 2 for values). Therefore, in order to bring the two values to an agreement, extinctions of about 3-4 visual magnitudes in the nuclear regions are required. These values are within the range of extinctions typical of  LIRGs (Alonso-Herrero et al. 2006).

\subsubsection {Ionized gas kinematics. Modeling the velocity field and measuring the dynamical mass}

The velocity field of the ionized gas exhibits a regular velocity field, typical of rotational motions. The major kinematic axis is well
 aligned with the stellar and ionized gas distributions traced by the optical continuum and  H$\alpha$ maps,
 respectively (see Figure 8). In addition, the velocity dispersion has its maximum value at the
 stellar center of the galaxy (within the angular resolution), as expected in galaxies dominated by pure rotation. Therefore, the observed velocity field has been modeled with the kinematic
  model
  described in equation (1) assuming that the radial and perpendicular velocity components are negligible ($\Pi(R,\theta) = Z(R,\theta) = 0$). 
  Considering an inclination value of 55 degrees (inferred from the stellar continuum map) the 908 data points were divided into 12 rings of spaxels located at increasing galactocentric distances and then equation (1) was solved with $\Omega$, $V_{sys}$, and $\theta_o$  as free parameters. The standard deviation was typically lower than 1 km s$^{-1}$ for most of the 12 independent fits performed, and only reached values of a few km s$^{-1}$ for the inner ( $<$ 0.4 kpc) and outer ($>$ 6.5 kpc) regions. Given that the values at large galactocentric distances were obtained using a limited range in azimuth they should be less reliable and are not considered further here.  The variation of the azimuthal velocity component, $\Omega$(R), is presented in Figure 9. This rotation curve reaches its maximun value ( 261 km s$^{-1}$) at a galactocentric distance of $\sim$ 3.5 kpc, then it keeps a relatively constant value of 254 $\pm$ 4 km s$^{-1}$ up to 6 kpc, when it starts to decline. The mean values for the systemic velocity ($V_{sys}$) and the PA of the major kinematic axis ($\theta_o$) evaluated at r $<$ 6.5 kpc were 5449 $\pm$ 11 km s$^{-1}$, and 115 $\pm$ 1 degrees (with an additional uncertainty of $\pm$ 180 degrees due to the uncertainty in $i$), respectively.  The results
    of the fit agree very well with previous values ($\Omega$= 262km s$^{-1}$, and $\theta_o$ = 294)
  obtained by Palunas \& Williams (2000) from Fabry-Perot H$\alpha$ maps. However, the  heliocentric
   velocity derived from the kinematic model disagrees with that given by Strauss et al (1992) who found 
    5543 $\pm$ 35 km s$^{-1}$, but shows a good coincidence with the value provided by  Sekiguchi et al.
    (1992) of 5436 $\pm$ 38 km s$^{-1}$. If the spiral arms seen in optical images of this galaxy  (e.g. in the DSS images) are trailing, the observed kinematics indicates anti-clockwise rotation, and the SW sector is the closest to us.

The observed and best-fit modeled velocity field together with the corresponding two-dimensional map velocity residuals (observed minus modeled) are presented in Figure 10. 
In general, the rotating disk model is a very good representation of the main motions of IRAS F12115$-$4656, with an average residual of  0.2 $\pm$ 12 km s$^{-1}$ (note that the contiguous E and W spaxels to the nucleus have relatively large residuals  of about 40 km s$^{-1}$ to the red and blue, respectively, due to the limited spatial resolution of the data).  
However, significant low amplitude ($\sim$ 20 -30  km s$^{-1}$) velocity residuals along PA 25 degrees  are identified (see Figure 10, right-upper panel and $\S$4.2.4 for a more detailed discussion).

    \begin{figure}
    \centering
   \includegraphics[angle=-0,width=9cm]{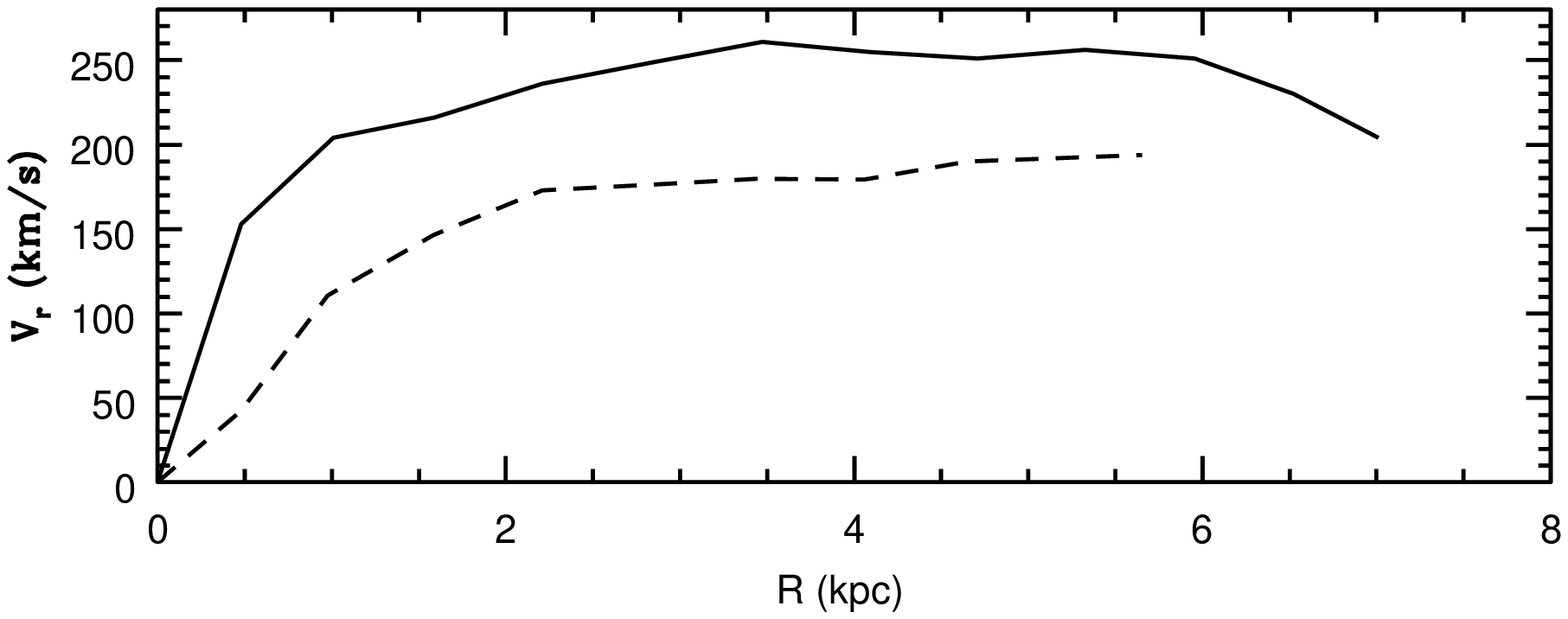}
      \caption{Rotation curves for the ionized (solid line) and neutral gas (dashed line). These curves represent the variation of the azimuthal component of the velocity field, $\Omega$, with the galactocentric distance, as inferred after fitting the data to a rotating disk model (see text). The nominal uncertainty (rms) in radial velocity at each of the evaluated radii is typically about 1 km s$^{-1}$, except for the  inner    ( $<$ 0.4 kpc) and outer ($>$ 6.5 kpc) regions, which is somewhat larger (3-5 km s$^{-1}$).}
         \label{FigVibStab}
   \end{figure}

   \begin{figure*}
    \centering
      \caption{{\bf  FULL VERSION at  http://damir.iem.csic.es/~aalonso/extragalactic/publications/publications.html} Figure presenting the results for the velocity fields of the ionized (top panels) and neutral (bottom panels) gas in IRAS F12115$-$4656. Observed (left panel), modelled (central panel) and residuals (observed minus modelled; right panel) are shown (see text for details).  The map corresponding to the neutral gas, which was inferred from the NaI lines, has been smoothed by averaging the spectra around each spaxel in order to increase the S/N. }
         \label{FigVibStab}
   \end{figure*}

The dynamical mass of the galaxy within a radius of 6.0 kpc corresponds to 8.7 $\times$ 10$^{10}$ M$_{\odot}$ for a derived rotation velocity of 251 $\pm$ 2 km s$^{-1}$ (see Figure 9). Note however, that if the velocity field of the neutral gas is considered (see next section), the dynamical mass within the same radius will be about  a factor of 2 lower. As for IRAS F06076$-$2139 (see $\S$4.1.7), a measurement of the total mass of the galaxy can be derived from the value of the central velocity dispersion and the effective radius. Since the VIMOS field-of-view is so small compared to the diameter of the galaxy,  the VIMOS continuum image cannot be used to establish the effective radius of the light distribution.  Instead, the effective radius measured in the ESO-Uppsala Galaxy Survey (Lauberts \& Valentijn 1989) in the B-band light is used. For an effective radius of 20.6 arcsec (7.7 kpc) and a central velocity dispersion of 111 km s$^{-1}$, a total mass of 1.65  $\times$ 10$^{11}$ M$_{\odot}$, equivalent to an 1.2 m$_*$ galaxy, is derived.

\subsubsection {Interstellar gas phases. Evidence for a kinematic and spatial decoupling of  the neutral and ionized gas components}

The physical properties of the cool neutral interstellar gas in galaxies can be obtained by measuring the characteristics of interstellar absorption lines such as the Na I doublet 5890, 5896 \AA\  in the optical (e.g. Martin, 2006; V\"{a}is\"{a}nen et al. 2007).  However, little is known so far about the two-dimensional kinematics of the  neutral gas phase in (U)LIRGs due to the difficulty of detecting these lines. Unlike the emission lines with large equivalent widths of typically 10 to 100 \AA, the NaI lines have  small equivalent widths of a few \AA\ . The high quality of the VIMOS data now allows us to measure the spatial distribution and kinematics of the neutral interstellar gas in two dimensions, making possible a direct one-to-one comparison with those derived for the warm ionized gas (see Figure 11). 

There are a number of remarkable features both in the spatial distribution and in the kinematics of the neutral gas that require further discussion.  The neutral gas distribution, traced by the equivalent width of the Na I 5890 $\AA$ line,  is similar to that of the stars with a clear peak at the photometric center of the galaxy, and extended emission along the same major axis as the stellar distribution (Figure 11). The ionized gas, on the other hand,
although sharing the same major axis, shows a very different spatial distribution dominated by the circumnuclear regions where the young ionizing star forming regions are located
(see $\S$4.2.1). Moreover, the fact that the equivalent width
of H$\alpha$ in the center of the galaxy is low, exactly where Na I has its maximum, could partly be due to an increase in the extinction towards the nuclear region where a larger concentration of cold dust associated with the cool neutral gas would exist.

\begin{figure*}
    \centering
      \caption{{\bf  FULL VERSION at  http://damir.iem.csic.es/~aalonso/extragalactic/publications/publications.html} Spatial distribution and kinematic properties of the neutral gas in IRAS F12115$-$4656 as traced by the  NaI doublet (top panels). The equivalent width of the
      H$\alpha$ line tracing the regions of ionized gas associated with young star-forming 
       regions is given for comparison (bottom left panel). The residuals of the velocity field
       and velocity dispersion of the neutral gas after subtraction of those measured for the 
       ionized gas are also given (central and right bottom panels) to highlight the differences
        in the kinematic properties of the two gas phases.}
         \label{FigVibStab}
   \end{figure*}

When compared with the kinematics of the ionized gas, both the velocity field and the velocity dispersion map of the neutral gas show distinct features (see Figure 11). On the one hand, the neutral gas shows indications of being affected by rotation as traced by the large peak-to-peak velocity amplitude (380 km s$^{-1}$) along a major axis centered in the nucleus of the galaxy. However, the velocity field shows some departures from a regular rotation pattern, and the maximum of the velocity dispersion is not at the dynamical center as expected in typical rotating systems. Moreover, the velocity dispersion of the neutral gas shows very high values ($\sigma$ $\sim$ 180 km$^{-1}$) at distances of about 0.8 kpc north and 2.7 kpc northwest of the nucleus, much higher than those measured in the ionized gas ($\sigma$ $\sim$ 30 to 40 km$^{-1}$) at the same locations and with no clear correspondence with any of the circumnuclear star-forming regions. Broad Na I lines such as those identified in the northwest regions of IRAS F12115$-$4656 have also been detected in some ULIRGs outside the nucleus (Martin 2006). 
 
To further compare the neutral and ionized gas kinematics, the rotational velocity pattern of the neutral gas has been substracted from that of the ionized gas. This map  (Figure 11, central lower panel) shows the morphology of a bipolar flow with velocities of up to $\pm$ 100 km s$^{-1}$, and with its axis oriented close to the major kinematic axis of the ionized gas. This behaviour can be explained by  the neutral gas rotating slower than the ionized gas but along a similar major kinematic axis. This scenario is confirmed when modeling the neutral gas velocity field with a rotating disk structure in a similar manner as for the ionized gas (see Figure 10 lower panels) .  As can be seen in these panels the model is also a good representation of the observed velocity field, with average residuals  of 0.1 $\pm$ 17 km s$^{-1}$. The average systemic velocity at distances of less than 5.5 kpc is 5460 $\pm$ 18 km s$^{-1}$ , in agreement with that of the ionized gas within the uncertainties. However, the rotation curve of the neutral gas (see Figure 9) indicates a rotational velocity  systematically lower than that of the ionized gas by $\sim$  70 - 90 km s$^{-1}$. Also, the kinematic major axis is oriented along PA 98 $\pm$ 3 degrees, significantly misaligned with respect to that of the ionized gas by 17 $\pm$ 3 degrees. 

Regarding the velocity dispersions, the central values for both neutral (116 km s$^{-1}$ ) and ionized (111 km s$^{-1}$) gas agree well, as expected if these values trace the mass in the nuclear regions. However, outside the nucleus, the overall residuals indicate that the velocity dispersion of the neutral gas is always higher than that of the ionized gas by 35 to
150 km s$^{-1}$ (see Figure 11, lower right panel for a two-dimensional map of the differences).  

The V/$\sigma$ ratio for the neutral and ionized gas shows even clearer differences. While the V/$\sigma$ values of the ionized gas are high (typically
$>$ 3) indicating a system dominated by rotation as in spirals, the corresponding V/$\sigma$ values for the neutral gas are lower (typically $<$ 1) suggesting the presence of a hotter system more affected by the random motions, like the stars in ellipticals. 

Summarizing, the kinematics of the neutral and ionized gas demonstrates that the dynamical status of the two gas phases is different, one (ionized gas) traces the presence of a rotating star-forming disk in the galaxy while the other (neutral gas) could be tracing the overall dynamics of the stellar distribution with some departures in specific regions.

\subsubsection {Evidence for non-rotational motion of the ionized and neutral gases: correlation with the velocity dispersion}

As already mentioned ($\S$4.2.2)  significant low amplitude ($\sim$ 20 -30  km s$^{-1}$) velocity residuals along $\sim$ PA -25 degrees have been identified (see Figure 10, right-upper panel).
Considering the kinematic model given by equation (1), such departures from pure rotation can be interpreted as radial motion in the plane of the disc (i.e., $\Pi(R,\theta) \neq $ 0 in eq. [1] ) and/or as motion perpendicular to the plane of the disc (i.e., Z(R,$\theta) \neq $ 0 in eq. [1] ). Purely radial motion  produces a pattern such that regions symmetric with respect to the nucleus have opposite signs in their velocities. This is not observed, thus favoring the hypothesis that they mainly represent motions perpendicular to the plane. Furthermore, taking into account the geometry of the disc, with the SW sector closest to us, the residuals suggest motions outwards the disc.  Interestingly, these regions with non-rotational motions do spatially correlate with regions of increased velocity dispersion (see right panel in Figure 8), indicating a physical connection between the two kinematic tracers.

The same kind of behaviour, and even clearly detected, is found in the neutral gas. The regions where the largest departures  from pure rotation are the detected (up to 50 km  s$^{-1}$) also exhibit the largest velocity dispersions. In general the velocity residual map (Figure 10 lower-right) has a remarkable spatial correlation with the velocity dispersion derived from the NaI lines (Figure 11 upper-right). A correlation between the measured velocity dispersions and the presence of non-rotational motion (mainly produced by tidal forces) has been reported at a higher interaction/merger scale in ULIRGs (Colina et al. 2005). Although IRAS F12115$-$4656 has only a weak interaction, we could be witnessing a low-scale effect of the same phenomenon.    

\section{Conclusions}

This paper presents the first results obtained from the integral field spectroscopic survey of local luminous
and ultraluminous infrared galaxies being carried out with the VIMOS instrument on the VLT.  This sample includes 42 local  (U)LIRGs systems (65 galaxies) and it is part of a larger representative sample of about 70 systems (z $<$ 0.26) also observed from the northern hemisphere, which covers the different morphologies from spirals to mergers and spans the entire luminosity range L$_{IR}$=10$^{11}-10^{12.6}$L$_{\odot}$.  The two galaxies analysed here (IRAS F06076$-$2139 and IRAS F12115$-$4656) have been selected as prototypes of the most common morphological types among (U)LIRGs, i.e.  interacting pairs and regular/weakly-interacting spirals, while also covering a wide range in IR luminosity. 

A) The VIMOS data of IRAS F06076$-$2139 clearly illustrate the potential of IFS in disentangling the complex ionizing and kinematic structure detected in pairs or mergers. The main conclusions for this system are:

a1) We have found that this system is formed by two independent low-intermediate mass galaxies, the G$_n$ (northern)  and G$_s$(southern), of 0.4 m$_*$  and 0.15 m$_*$, respectively, and  with a relative projected velocity of about 550 km s$^{-1}$. Therefore, it is unlikely that these galaxies will ever merge. G$_n$ is characterized by an AGN-like ionization, and large velocity dispersions ($\geq$ 100 km s$^{-1}$), while G$_s$ has HII characteristics and lower velocity dispersions  (by at least a factor of two)  than  G$_n$. 

a2) The morphology and velocity field of the ionized gas in G$_s$ is consistent with the presence of an expanding, rotating ring of about 20 kpc (deprojected major axis) in size. This expanding ring is interpreted as the result of a head-on collision with an intruder that took place some 1.4 $\times$ 10$^{8}$ years ago. The characteristics of G$_n$ rule out this galaxy  as the intruder. 

a3) A chain of H$\alpha$ emitting regions kinematically associated with G$_s$ at distances of about 12 kpc have been detected. These regions are  identified as young, tidally-induced star-forming regions (some may be candidate tidal dwarf galaxies) formed as a consequence of a past interaction.

B) The VIMOS data of IRAS F12115$-$4656 show the potential of IFS in the kinematic study of the different phases of the interstellar gas: the neutral gas traced by the NaI absorption doublet, and the ionized gas traced by the H$\alpha$ or [NII] emission lines. The main conclusions for this system are:

b1) The neutral and ionized gas morphology show a different structure indicating that they have a different spatial distribution. The neutral gas closely follows the stellar distribution while the ionized gas traces mostly the location of HII regions in the circumnuclear and external regions of the galaxy.

b2) The ionized gas shows typical kinematic characteristics of massive rotating disks,  with a large amplitude (peak-to-peak of 520 km s$^{-1}$), a maximum of the velocity dispersion at the kinematic/light centre, and a major kinematic axis coincident with the stellar photometric axis. A dynamical mass of 8.7 $\times$ 10$^{10}$ M$_{\odot}$ is measured inside a region of 6 kpc radius. The neutral gas, although consistent  on a large scale with a rotating system, shows clear departures from pure rotation and it is dynamically hotter than the ionized gas, being more affected by random motions.

b3) Despite being dominated by rotation the ionized gas shows large-scale (10 kpc),  low-amplitude (20-30 km s$^{-1}$) non-rotational motion, which spatially correlates  with regions of increased velocity dispersion.  The neutral gas also exhibits an even clearer spatial correlation between departures from rotation and high velocity dispersion values.   

\begin{acknowledgements}
Based on observations carried out at the European Southern Observatory, Paranal (Chile), programs 076.B-0479(A) and 078.B-0072(A). Also based on observations made with the Hubble Space Telescope obtained from the ESO/ST-ECF Science Archive Facility. This work has been supported by the Spanish Ministry for Education and Science under grant PNE2005-01480 and ESP2007-65475-C02-01. AMI also acknowledges the support under grant AYA2004-08260-C03-01. This research has made use of the NASA/IPAC Extragalactic Database (NED) which is operated by the Jet Propulsion Laboratory, California Institute of Technology, under contract with the National Aeronautics and Space Administration. The authors thank the anonymous referee.
      
\end{acknowledgements}

\begin{table*}[t]
\caption{VIMOS Luminous and Ultraluminous Infrared Galaxy sample} 
\label{LIRGs}
\begin{minipage}{2\columnwidth}
\centering
\resizebox{1\textwidth}{!}{
\begin{tabular}{ccccccccccl}
\hline
\hline

ID1 & ID2 &$\alpha$& $\delta$& $z$\footnote{As given in the NASA  
Extragalactic
Database (NED)}
& $v$
&$D$\footnote{Luminosity distances using the calculator by Wright (2006) for a lambda-cosmology with H$_0
$=70 kms$^
{-1}$Mpc$^{-1}$, $\Omega$$_M$=0.3, and $\Omega_{\Lambda}$=0.7. }
 &scale
& $\log L_{\mathrm{ir}}$\footnote{Logarithm of the infrared luminosity,
L$_{ir}$=L(8-1000$\mu$m), in units of solar bolometric luminosity,  
calculated
using the fluxes in the four IRAS bands as given in Sanders et al. (2003), when available. Otherwise from the IRAS fluxes given in IRAS Point Source and Faint Source catalogues (Moshir et al. 1990).}
& Class\footnote{Morphology class according to a simplified  
classification
scheme similar to that given by Veilleux et al. (2002) for ULIRGs.  
Note that
some objects are just single isolated objects without traces of undergoing
a merging process, and they are classified as class 0. Galaxy mergers  
in a
pre-coalencense phase are classified as class 1, and objects with  
peculiar
morphologies suggesting a post-coalescence merging phase are  
classified as class
2. Note that this classification is based on existing imaging data and in some cases it may change after complete reduction and analysis of the sample.}
& Comments\\
IRAS & Other &(J2000)& (J2000)&  & (km/s) & (Mpc) & (pc/arcsec) & (L$_
\odot$) & & \\
\hline
F01159$-$4443 &   ESO~244$-$G012    &01:18:08.1 & -44:27:51 &  0.022903& 6866  &
99.8 &   462  & 11.48 & 1 &
 Double system (dist.$\sim$17$''$)\\
F01199$-$2307$\dagger$ &                     &01:22:20.9 & -22:52:07 & 0.156000 & 46768 &
744.5 &  2701 & 12.41 & 2 &
 Kim et al. (2002) /Irregular with prominent tidal tails in  
WFPC2 image \\
           & ESO-297-G011 / G012 &01:36:23.4 & -37:19:18 & 0.017305 & 5188  & 
75.0 &   352 & 11.18 & 1 & 
NGC~0633. Double system (dist. $\sim$ 1')\\
F02021$-$2104 &                     &02:04:27.3 & -20:49:41 & 0.116000 & 34776 &
539.6  &  2101 & 12.14 & 0 \footnote{This system was classified by Veilleux et al. (2002) as a late-stage merger. However, we consider that its morphology can be simply interpreted as a nearly edge-on disc galaxy.} & 
Kim et al. (2002) / Edge-on spiral in WFPC2 ?\\
F02587$-$6336$\dagger$ &                     &02:59:43.4 & -63:25:00 & 0.264800 & 79385 &
1345.5 &  4078 & 12.92 & 1 &
 Compact  Ir in WFPC2 image\\
F04315$-$0840 &  NGC 1614           &04:33:59.8 & -08:34:44 & 0.015938 & 4778  &
69.1   & 325   & 11.69 & 2 & 
Arp~186. Spiral with large scale tidal structures \\
F05020$-$2941$\dagger$ &   Abell  3307       &05:04:00.7 & -29:36:55 & 0.154000 & 46168 &
734.0  & 2672  & 12.43 & 2 &
Distorted source with short tidal tails according to Veilleux et al. (2002)\\
F05120$-$4811$\dagger$ &                     &05:13:23.5 & -48:07:57 & 0.162699 & 48776 &
779.7  & 2796  & 12.58 & 2 &
 Compact Ir in WFPC2 image\\
F05189$-$2524 &                   &05:21:01.5 & -25:21:45 & 0.042563 & 12760 &
188.2  & 839   & 12.19 & 2 &
 Tidal structures in the ESO-Danish-1.54m images \\
F06035$-$7102 &                     &06:02:54.0 & -71:03:10 & 0.079465 & 23823 &
360.7 & 1501  & 12.26 & 1 & 
Double system (dist. $\sim$ 10$''$) in WFC2 image\\
F06076$-$2139 &                   &06:09:45.8 & -21:40:24 & 0.037446 & 11226 &
165.0  & 743   & 11.67 & 1 &
 On-axis merger\\
F06206$-$6315 &                     &06:21:01.2 & -63:17:23 & 0.092441 & 27713 &
423.3  & 1720 & 12.27 & 2 &
Irregular with a prominent arm / tidal tail in WFPC2  image \\
F06259$-$4708& ESO~255$-$IG 007   &06:27:22.0 & -47:10:54 & 0.038790 & 11629 &
171.1  & 769   & 11.91 & 1 & 
Triple  System\\
F06295$-$1735 &  ESO 557$-$G002     &06:31:47.2 & -17:37:17 & 0.021298 & 6385  &
92.7 &  431  & 11.27 & 0 & 
Spiral \\
F06592$-$6313 &                   &06:59:40.2 & -63:17:53 & 0.022956 & 6882  &
100.0  & 464  & 11.22 & 0 & 
Spiral (Ir), wide pair\\
F07027$-$6011& AM~0702$-$601      &07:03:24.1 & -60:15:23 & 0.031322 & 9390  &
137.4  & 626 & 11.64 & 0 & 
Spiral \\

F07160$-$6215& NGC~2369           & 07:16:37.7& -62:20:37 & 0.010807 & 3240 & 
46.7 & 221   &11.16 & 0  &
 Spiral with irregular inner parts\\
F08355$-$4944 &                   &08:37:01.8 & -49:54:30 & 0.025898 & 7764 &
113.1 & 521   &11.60 & 2  & 
Irregular morphology in ACS image \\
F08424$-$3130 & ESO~432$-$IG006   &08:44:27.6 & -31:41:50 & 0.016165 & 4846 &
70.1  & 329   &11.04 & 1  & 
Double system (off-set 25'') \\
F08520$-$6850 & ESO~60$-$IG016    &08:52:29.9 & -69:01:58 & 0.046315 & 13885&
205.4   & 909   & 11.83& 1  & 
Secondary nucleus at $\sim$ 2$''$. \\
F09022$-$3615 &                   &09:04:12.7 & -36:27:01 & 0.059641 & 17880&
267.0   & 1153  & 12.32& 2  &
 Faint tidal tail in DSS image ? \\
F09437+0317 & IC-563/ 564       &09:46:20.6 &+03:03:30  &0.020467 & 6136 & 89.0 
& 415    & 11.21& 1  &
Arp~303. Double system (dist. $\sim$ 90$''$)\\
F10015$-$0614 & NGC-3110          &10:04:02.1 & -06:28:29 & 0.016858 & 5054 &
73.1  & 343   & 11.31& 0  &
 Spiral. Interacting with galaxy at SW ? \\
F10038$-$3338 & IC2545           &10:06:35.0 & -33:51:30 & 0.034100 & 10223&
149.9   & 679   & 11.77& 2  &
Prominent spiral arm or tidal tail in ACS image \\
F10173+0828 &                     &10:20:00.2 & +08:13:34 & 0.049087 & 14716&
218.1   & 961   & 11.86& 2  &
 Linear structure in ACS image with evidence of a faint tilted tidal tail\\
F10257$-$4338 & NGC~3256          & 10:27:51.3& -43:54:14 & 0.009354 & 2804 &
40.4  & 192   & 11.69& 2  &
 Long tidal tails. Only innner parts in FoV \\
F10409$-$4556 & ESO~264$-$G036   & 10:43:07.7&  -46:12:45& 0.021011 & 6299 &
91.4  & 425   & 11.26& 0  &
 Spiral\\
F10485$-$1447$\dagger$    &                  &10:51:03.1 &  -15:03:22& 0.133000 & 39872&
625.6   & 2363  & 12.32& 1  & 
Kim et al. (2002). Double/triple  system\\
F10567$-$4310 & ESO~264$-$G057    & 10:59:01.8& - 43:26:26& 0.017199 & 5156 &
74.6  & 350   & 11.07& 0  &
 Spiral\\
F11254$-$4120 & ESO~319$-$G022    & 11:27:54.1& -41:36:52 & 0.016351 & 4902 &
70.9  & 333  & 11.04 & 0  &
 Spiral. Inner ring. Barred\\
F11506$-$3851 & ESO~320$-$G030    & 11:53:11.7& -39:07:49 & 0.010781 & 3232 &
46.6  & 221  & 11.30 & 0  &
 Spiral\\
F12042$-$3140 & ESO~440$-$IG 058  & 12:06:51.9& -31:56:54 & 0.023203 & 6956 &
101.1   & 468 & 11.37 & 1  &
 Double system. Tidal structures\\
F12115$-$4656 & ESO~267$-$G030   & 12:14:12.8& -47:13:43 & 0.018489 & 5543 &
80.3  & 375  & 11.11 & 0  &
 Spiral, wide pair\\
F12116$-$5615 &                   & 12:14:22.1& -56:32:33 & 0.027102 & 8125 &
118.5   & 545  & 11.61 & 2  &
Inner spiral-like structure and outer  elliptical morphology in ACS image.\\
F12596$-$1529$\star$ & MCG$-$02$-$33$-$098 & 13:02:19.7& $-$15:46:03 & 0.015921& 4773 &
69.0& 324 & 11.07 & 2  &
 Sc pec. Close interacting pair  \\
F13001$-$2339 & ESO~507$-$G070    & 13:02:52.3& -23:55:18 & 0.021702 & 6506 &
94.5  & 439  & 11.48 & 2  &
 Irregular. Tidal structures\\
F13229$-$2934 $\star$ & NGC 5135         & 13:25:44.0&$-$29:50:01& 0.013693 & 4105 &
59.3  & 280  & 11.29 & 0  &
 SB spiral. In group \\
F14544$-$4255 $\star$ & IC 4518         & 14:57:42.9& $-$43:07:54& 0.015728& 4715 &
68.2 &  320  & 11.11 & 1  &
 Sc pec. close interacting pair \\
F17138$-$1017 $\star$ &                  & 17:16:35.8& $-$10:20:39& 0.017335& 5197 &
75.2 &  352  &11.41  & 2  &
 Irregular \\
F18093$-$5744 $\star$ & IC 4687/4686     & 18:13:39.6& $-$57:43:31& 0.017345& 5200 &
75.3 &  353  &11.57  & 1  &
 Spiral pec. Close interacting pair \\
F21453$-$3511$\star$  & NGC 7130         & 21:48:19.5& $-$34:57:05& 0.016151& 4842 &
70.0 &  329  & 11.41 & 2  &
 Sa pec ? \\
F22132$-$3705$\star$ & IC 5179          & 22:16:09.1& $-$36:50:37& 0.011415& 3422 &
49.3 &  234  & 11.22 & 0  &
 SA(rs)bc, isolated \\

\hline
\hline

\end{tabular}
}
\end{minipage}

$\dagger$ At the redshift of this object the standard instrumental setting was such 
 that  it excluded the H$\alpha$ region, but it includes the H$\beta$ and the [OIII] lines.  $\star$ Galaxies with pending observations.
\end{table*}

\begin{table*}
\begin{minipage}{2\columnwidth}
\caption{Main physical properties of IRAS F06076$-$2139 and IRAS F12115
$-$4656 derived from VIMOS IFS}             
\label{VIMOS}      
\centering                          
\renewcommand{\footnoterule}{}  
\begin{tabular}{cccc}        
\hline\hline                 
Quantity& IRAS F06076$-$2139 & IRAS F06076$-$2139 & IRAS F12115$-$4656\\
& G$_n$ & G$_s$ &  \\
\hline
V$_{sys, H\alpha}$\footnote{systemic velocity in km s$^{-1}$ as obtained from the best  
fit model to the observed H$\alpha$ velocity field. The errors include both wavelength calibration and model fitting uncertainties} & 11236 $\pm$ 10 & 11750 $\pm$ 10 & 5449 $\pm$ 11\\
V$_{nucleus, H\alpha}$\footnote{observed systemic velocity in km s$^{-1}$ as obtained from the nuclear H$\alpha$ line} & 11186 $\pm$ 25 & 11737 $\pm$ 25 & 5453 $\pm$  
15\\
V$_{sys, NaD}$\footnote{systemic velocity in km s$^{-1}$ as obtained from the best fit  
model to the observed NaD velocity field in IRAS F12115$-$4656. Uncertainties include both wavelength calibration and model fitting} & $-$ & $-$ & 5460 $\pm$ 18 \\
V$_{nucleus, NaD}$\footnote{observed systemic velocity in km s$^{-1}$ as obtained from the  nuclear NaD line} & $-$ & $-$  & 5446 $\pm$  15\\
$\Omega_{H\alpha}$\footnote{rotational velocity in km s$^{-1}$ as  
obtained from best fit model to the observed H$\alpha$ velocity field} & 76 $\pm
$ 5 & 61 $\pm$ 3 & 251 $\pm$ 2\\
$\sigma_{nucleus}$\footnote{velocity dispersion in km s$^{-1}$ measured using the 
nuclear H$\alpha$ line profile} & 104 $\pm$ 10 & 63 $\pm$ 10& 111 $\pm$ 10\\
R$\footnote{galactocentric radius in kpc for the rotational velocity given above} $& 5.6 & 10.1 & 6.0 \\
R$_{e}$\footnote{effective radius in kpc from the available ACS F814W image for IRAS  
F06076$-$2139, and from Lauberts \& Valentijn 1989 for IRAS F12115$-
$4656} & 2.76 & 2.73 & 7.7 \\
M$_{dyn,\Omega}$\footnote{dynamical mass in 10$^{10}$M$_{\odot}$ as derived from the  
rotational velocity and galactocentric radius given above} & 0.75 &  
1.7 & 8.7  \\
M$_{dyn,\sigma}$\footnote{dynamical mass in 10$^{10}$M$_{\odot}$ derived from the  
effective radius and velocity dispersion values given above. The  
values in parenthesis are given in units of m$_*$
(1.4 $\times$ 10$^{11}$M$_{\odot}$, Cole et al. 2001)} & 5.2 (0.4) &  
1.9 (0.15) & 16.5 (1.2) \\
F$_{H\alpha}$\footnote{observed flux in 10$^{-14}$ erg s$^{-1}$ cm$^{-2}$ not  
corrected for internal extinction} & 1.71 & 1.26 & 31.3\\
L$_{H\alpha}$\footnote{in 10$^{41}$ erg s$^{-1}$ from observed H$
\alpha$ fluxes given above} & 0.56 & 0.41 &  2.42\\
L$_{IR}$\footnote{in 10$^{11}$ L$_{\odot}$} & 4.68 & $-$ & 1.29 \\
SFR$_{H\alpha}$\footnote{Star formation rate in M$_{\odot}$ yr$^{-1}$ assuming SFR = 7.9 $
\times$ 10$^{-42}$ $\times$ L$_{H\alpha}$} & 0.4 & 0.3 & 1.9 \\
SFR$_{IR}$\footnote{Star formation rate in M$_{\odot}$ yr$^{-1}$ assuming SFR = 4.5 $
\times$ 10$^{-44}$ $\times$ L$_{IR}$ (Kennicutt 1998). The value for IRAS F06076$-$2139  
represents the contribution from the two galaxies, G$_n$ and G$_s$} &  
81 & $-$ & 22 \\
log([OI]/H$\alpha$)\footnote{not corrected for internal extinction,  
ratios represent the values obtained for the nucleus} & $-$1.42 
& $-$1.48 & $-$1.31\\
log([NII]/H$\alpha$) $^p$& $-$0.12  
& $-$0.29  &  $-$0.15\\
log([SII]/H$\alpha$) $^p$& $-$0.50  
& $-$0.46 & $-$0.39\\
\hline                                   
\end{tabular}
\end{minipage}
\end{table*}



\begin{thebibliography}{}

\bibitem[2006]{Adel06} Adelman-McCarthy, J. et al. 2006, ApJS, 162, 38
\bibitem[2006]{Alon06} Alonso-Herrero, A., et al. 2006, ApJ, 650, 835
\bibitem[1996]{Arr96} Arribas, S., Mediavilla, E., \& Garc\'\i a-Lorenzo, B. 1996, ApJ, 463, 509
\bibitem[2006]{Arr98} Arribas, S. et al. 1998, SPIE, 3355, 821
\bibitem[1982]{Arp82} Arp, H.C., \& Madore, B.F. 1982, J.R. Astron. Soc. Can.76, 315
\bibitem[2006]{Bec06} Beckwith, S.V.W. et al. 2006, AJ, 132, 1729
\bibitem[1999]{Ber99} Bertola, F., \& Corsini, E.M. 1999, in "Galaxy Interactions at Low and High Redshift", Proceedings of IAU Symposium 186. Edited by J. E. Barnes, \& D. B. Sanders, p.149
\bibitem[1996]{bu96} Buta, R. 1996, AJ, 111, 591
\bibitem[2007]{Cap07} Capak et al. 2007, ApJS, 172, 99
\bibitem[2001]{Cole01} Cole, S. et al. 2001, MNRAS, 326, 255
\bibitem[2005]{Col05} Colina, L., Arribas, S., \& Monreal-Ibero, A. 2005, ApJ, 621, 725
\bibitem[2001]{Col01} Colless et al. 2001, MNRAS, 328, 1039
\bibitem[2006]{Das06} Dasyra, K.M. et al. 2006, ApJ, 651, 835 
\bibitem[2001]{Col05} del Burgo, C., Peletier, R.F., Vazdekis, A., Arribas, S., \& Mediavilla, E., 2001, MNRAS, 321, 227
\bibitem[2002]{deZ02} de Zeeuw, P.T. et al. 2002, MNRAS, 329, 513
\bibitem[2006]{El06} Elmegreen, D.M., \& Elmegreen, B.G. 2006, ApJ, 651, 676
\bibitem[2006]{FS06} Flores, H., Hammer, F., Puech, M., Amram, P., \& Balkowski, C. 2006, A\&A, 455, 107
\bibitem[2006]{FS06} Forster-Schreiber, N.M. et al. 2006, ApJ, 645, 1062
\bibitem[2004]{Fra04} Frayer, D.T., Reddy, N.A., Armus, L., Blain, A.W., Scoville, N., \& Smail, I. 2004, AJ, 127, 728
\bibitem[2007]{GM07} Garc\'\i a-Mar\'\i n 2007 PhD Thesis, Universidad Autonoma de Madrid 
\bibitem[2006]{Gen98} Genzel, R. et al. 1998, ApJ, 498, 579
\bibitem[2006]{Gen98} Genzel, R., Tacconi, L.J., Rigopoulou, D., Lutz, D., \& Tecza, M. 2001, ApJ, 563, 527
\bibitem[2004]{Gia04} Giavalisco, M. et al. 2004, ApJ, 600, L93
\bibitem[1988]{how} Howarth, J. D., \& Murray, J. 1988, Starlink User Note 50
\bibitem[2001]{ka01} Kannappan, S.J., Fabricant, D.G. 2001, AJ, 121, 140
\bibitem[2006]{kew06} Kewley, L.J., Groves, B., Kauffmann, G., \& Heckman, T. 2006, MNRAS, 372, 961
\bibitem[2002]{kim02} Kim, D.-C., Veilleux, S., \& Sanders, D.B. 2002, ApJS, 143, 277
\bibitem[2007]{law07} Law, D.R., Steidel, C.C., Erb, D., Larkin, J.E., Pettini, M., Shapley, A.E., \& Wright, S.A. 2007, astro-ph0707.3634
\bibitem[2007]{la89} Lauberts, A. \& Valentijn, E.A. 1989, The Surface Photometry Catalogue of the ESO-Uppsala Galaxies
\bibitem[LeF$\grave{e}$vre et al.(2003)]{lef03} LeF$\grave{e}$vre, O., et al. 2003, \procspie, 4841, 1670
\bibitem[2006]{Ma06} Martin, C.L. 2006, ApJ, 647, 222
\bibitem[1997]{Med97} Mediavilla, E, Arribas, S., Garc\'\i a-Lorenzo, B., \& del Burgo, C. 1997, ApJ, 488, 682
\bibitem[1981]{Mih81} Mihalas, D., \& Binney, J. 1981, {\it Galactic Astronomy. Structure and Kinematics} Publisher: W.H. Freeman and Co.
\bibitem[2006]{Mon06} Monreal-Ibero, A., Arribas, S., \& Colina, L. 2006, ApJ, 637, 138
\bibitem[2007]{Mon07} Monreal-Ibero, A., Colina, L., Arribas, S., \& Garc\'\i a-Mar\'\i n, M. 2007, A\&A, 472, 421
\bibitem[1989]{Mo89} Moshir, M. et al. 1990, IRAS Faint Source Catalogue, version 2.0
\bibitem[1996]{os96} Osterbrock, D.E., Fulbright, J.P., Martel, A.R., Keane, M.J., Trager, S.C., \& Basri, G. 1996, PASP, 108, 277
\bibitem[2000]{Pal00} Palunas, P., \& Williams, T.B. 2000, AJ, 120, 2884 
\bibitem[1999]{Pel99} Peletier, R. et al. 1999, MNRAS, 310, 863 
\bibitem[2007]{Pel07} Peletier, R. et al. 2007, MNRAS, 379, 445
\bibitem[2005]{Pe05} P\'erez-Gonz\'alez, P.G. et al. 2005, ApJ, 630, 82
\bibitem[2007]{Pue07} Puech, M., Hammer, F., Lehnert, M.D., \& Flores, H. 2007, A\&A, 466, 83 
\bibitem[2006]{Rix06} Rix, H.-W. et al. 2004, ApJS, 152, 163 
\bibitem[2005]{Ro05} Roth, M.M. et al. 2005, PASP, 117, 620
\bibitem[1996]{Sand96} Sanders, D.B., \& Mirabel, I.F. 1996, ARAA, 34, 749
\bibitem[2003]{Sand03} Sanders, D.B., Mazzarella, J.M., Kim, D.-C., Surace, J.A., \& Soifer, B.T. 2003, AJ, 126, 1607
\bibitem[2006]{Sar06} Sarzi, M. et al. 2006, MNRAS, 366, 1151
\bibitem[1992]{Sek92} Sekiguchi, K.S., \& Wolstencroft, R.D. 1992, MNRAS, 255, 581
\bibitem[1993]{sch93} Schommer, R.A., Bothun, G.D., Williams, T.B., \& Mould, J.R. 1993, AJ, 105, 97
\bibitem[1992]{Sco00} Scoville, N.Z., Evans, A.S., Thompson, R., Rieke, M., Hines, D.C., Low, F.J., Dinshaw, N., Surace, J., \& Armus, L. 2000, AJ, 119, 991
\bibitem[2006]{Skru06} Skrutskie et al. 2006, AJ, 131, 1163
\bibitem[2004]{Smi04} Smith, J.K., Bunker, A.J., Vogt, N.P., Abraham, R.G., Arag\'on-Salamanca, A., Bower, R.G., Parry, I.R., Sharp, R.S., \& Swinbank, A.M. 2004, MNRAS, 354, L19
\bibitem[1992]{stra92} Strauss, M.A., Huchra, J.P., Davis, M., Yahil, A., Fisher, K.B., \& Tonry, J. 1992, ApJS, 83, 29
\bibitem[1987]{Stru87} Struck-Marcell, C., \& Appleton, P.N. 1987, ApJ, 323, 480
\bibitem[2007]{Vaisanen07} V\"{a}is\"{a}nen, P. et al. 2007, MNRAS (in press, astro-ph 0708.2365)
\bibitem[2002]{Vei02} Veilleux, S., Kim, D.-C., \& Sanders, D.B. 2002, ApJS, 143, 315
\bibitem[2006]{Wr06} Wright, E.L. 2006, PASP, 118, 1711

\end{thebibliography}
\end{document}